\documentclass[twocolumn,showpacs,preprintnumbers,amsmath,amssymb,aps,dvips,a4paper]{revtex4}
\usepackage{amsmath}
\usepackage[dvips]{graphicx}
\usepackage{dcolumn}
\newcolumntype{d}[1]{D{.}{.}{#1}}
\usepackage{bigdelim}
\usepackage{bm}
\usepackage{array}
\usepackage{longtable}
\usepackage{lscape}
\usepackage{float}
\usepackage{rotating}
\usepackage{multirow}
\usepackage{setspace}

\makeatletter

\newcommand{\Rmnum}[1]{\expandafter\@slowromancap\romannumeral #1@}
\makeatother

\begin{document}

\preprint{APS/123-QED}

\title{Quantum Dimensional Transition in Spin-$\frac{1}{2}$ Antiferromagnetic Heisenberg Model on A Square Lattice and Space Reduction in Matrix Product State}
\author{Lihua Wang$^{1}$}
\email{wanglihua94@tsinghua.org.cn}
\author{Kwang S. Kim$^{1}$}
\email{kim@unist.ac.kr}
\affiliation{$^1$
Department of Chemistry, School of Natural Science, Center for Superfunctional Materials, Ulsan National Institute of Science and Technology (UNIST), Ulsan 44919, Republic of Korea}

\date{\today}

\begin{abstract} 
We study the spin-$\frac{1}{2}$ antiferromagnetic Heisenberg model on an infinity-by-$N$ square lattice for even $N$'s up to $14$. Previously, the nonlinear sigma model perturbatively predicts that its spin rotational symmetry asymptotically breaks when $N\rightarrow \infty$, i.e., when it is two-dimensional (2D). However, we identified a critical width $N_c = 10$ for which this symmetry breaks spontaneously. It defines a dimensional transition from one-dimension (1D) including quasi-1D to 2D. The finite-size effect differs from that of the $N$-by-$N$ lattice. The ground state (GS) energy per site approaches the thermodynamic limit value, in agreement with the previously accepted value, by one order of $1/N$ faster than when using $N$-by-$N$ lattices in the literature. We build and variationally solve a matrix product state (MPS) on a chain, converting the $N$ sites in the rung into an effective site. We show that the area law of entanglement entropy does not apply when $N$ increases in our method, and show that the reduced density matrix of each effective site will have a saturating number of dominant diagonal elements with increasing $N$. These two characteristics make the MPS rank needed to obtain a demanded energy accuracy quickly saturate when $N$ is large, making our algorithm efficient for large $N$'s. And, the latter enables space reduction in MPS. Within the framework of MPS, we prove a theorem that the spin-spin correlation at infinite separation is the square of staggered magnetization and demonstrate that the eigenvalue structure of a building MPS unit of $\langle g\mid g\rangle$, $\mid g\rangle$ being the GS, is responsible for order, disorder and quasi-long-range order. 
\end{abstract}

\maketitle

\section{\label{sec:introductionl}Introduction}
\subsection{\label{subsec:description}Problem Description}

With the advancement of experimental probes on the quantum spin system in both one-dimension (1D)\cite{Jompol2009} and two-dimension (2D)\cite{Gross2017}, rich ground state (GS) phases such as the disordered Tomonaga-Luttinger spin liquid and the ordered non-collinear antiferromagnetic state are revealed. An ideal model describing them is the antiferromagnetic spin-$\frac{1}{2}$ Heisenberg model. Such a model on an infinite 1D lattice is not ordered with power-law-decaying spin-spin correlations\cite{Bethe1931}. But, "more is different"\cite{Anderson1972}. When a collection of infinite 1D lattices is isotropically coupled to form an infinity-by-$N$ square lattice called spin ladder, the GS is predicted to be not ordered with exponentially-decaying correlations in a few quasi-1D lattices\cite{Greven1996} but ordered in 2D\cite{Manousakis1991}. Hereafter, we assume an infinity-by-infinity square lattice in the thermodynamic limit in our reference to a 2D lattice and confine the discussion within even $N$'s. It implies that there is at least one dimensional transition from 1D to 2D either asymptotically at $N=\infty$, as the nonlinear sigma model (NLSM) predicts\cite{Chakravarty1996,Sierra1996}, or critically at some finite width $N_c$. Note that this change of dimensional characteristics occurs purely due to a critical change of lattice topology, different from those caused by the variation of temperature or spin-spin coupling anisotropy\cite{Hoang1977,Kung2017,Raczkowski2013,Disseler2015}. Owing to the perturbative nature of mapping the spin-$\frac{1}{2}$ ladder to NLSM, its prediction that the gap exponentially decays with increasing $N$ implies the existence of some threshold beyond which the perturbation would be inapplicable. In fact, its prediction on the existence of gap was numerically checked only for $M$-by-$N$ lattices with $M\gg N$ and $N$ up to $6$\cite{White1994,Dagotto1996}. And, the latest size-scaling of $N$-by-$N$ or $M$-by-$N$ lattices in the literature did not handle larger $N$ yet and hence did not capture any dimensional transition\cite{Stoudenmire2012,Ramos2014} though the possibility is not excluded\cite{Landsman2013}. Therefore, it is worth exploring the possibility of such a quantum dimensional transition at finite $N$ for a true infinity-by-$N$ lattice of larger $N$'s, by monitoring the emerging order parameter such as the stagger magnetization and the spin-spin correlation at infinite separation.  

However, this is not an easy task for both analytic methods and numerical methods. Other than the aforementioned NLSM, on the analytic side, Bethe ansatz\cite{Bethe1931} only works for $N=1$; Bosonization\cite{Luther1974,Luther1975} predicts a power-law decay of the spin-spin correlation $C\left(r\right)\equiv\langle S^z_{\left(i,1\right)} S^z_{\left(i+r,1\right)}\rangle=r^{-1}$ for $N=1$, r being the spin-spin separation; Conformal field theory (CFT)\cite{Affleck1990,Affleck1998a} further predicts a logarithmic correction multiplying with the power function, which was confirmed\cite{Wang2012} to be asymptotically effective after $1000$ lattice separations; CFT\cite{Shelton1996} also gives a solution in limiting cases for $N>1$ such as the 2D Ornstein-Zernike form of spin-spin correlations for weakly coupled spin ladders; the spin wave theory (SWT)\cite{Chernyshev2009,wang2011} essentially provides an approximation in the continuum limit and assumes the magnon excitation, excluding the spinon excitation, hence giving no decisive observation of the quantum dimension transition. Numerically, it is not feasible for statistical methods such as Monte Carlo method\cite{Ceperley1980} which otherwise is powerful in searching energy of a finite system. Finite $M$-by-$N$ lattices, with $M\gg N$, were simulated\cite{Greven1996}, trying to scale away the finite size effect. A similar finite lattice was also simulated\cite{Ramos2014} by the density matrix renormalization group method (DMRG)\cite{White1992,White1993}. But sweeping an infinity-by-$N$ lattice to establish long-range spin-spin correlations for large $N$ is not yet practical. Variants of DMRG such as infinite time evolving block decimation (iTEBD) method were applied to the case of $N=2$\cite{Furukawa2010}, yielding results conflicting with that by the infinite quasi-1D entanglement perturbation theory (iqEPT)\cite{Wang2015}, and so on\cite{Kariyado2015}, but not for larger $N$ because of the rapid increase of the number of density matrix elements needed for a sufficient accuracy. Tensor network state (TNS)\cite{Nishino2000,Nishio2004} based methods such as the infinite projected entangled pair state (iPEPS)\cite{Jordan2008,Shi2016}, illustrated in Fig.\ref{fig:TNSandMPS} (a) and natively designed for an infinite 2D system, still do not show sufficient efficiency to tackle the tensor's bond index size greater than a few dozens\cite{Shi2016,Ehlers2017}. It hinders its application to investigate the very fine structure of the spin-spin correlation covering large separations within large systems\cite{Li2012}. 

\begin{figure}
	\begin{center}
		$\begin{array}{ccc}
		&\mbox{(a) original lattice} 	& \mbox{(b) effective lattice} \\
		& \includegraphics[width=10.5pc]{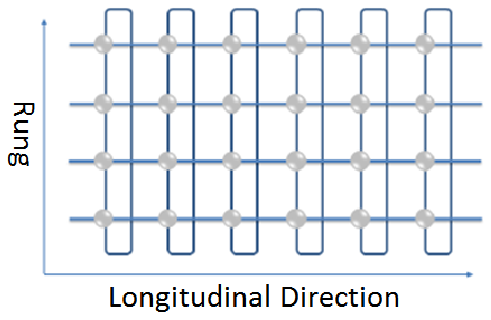}& \includegraphics[width=10.5pc]{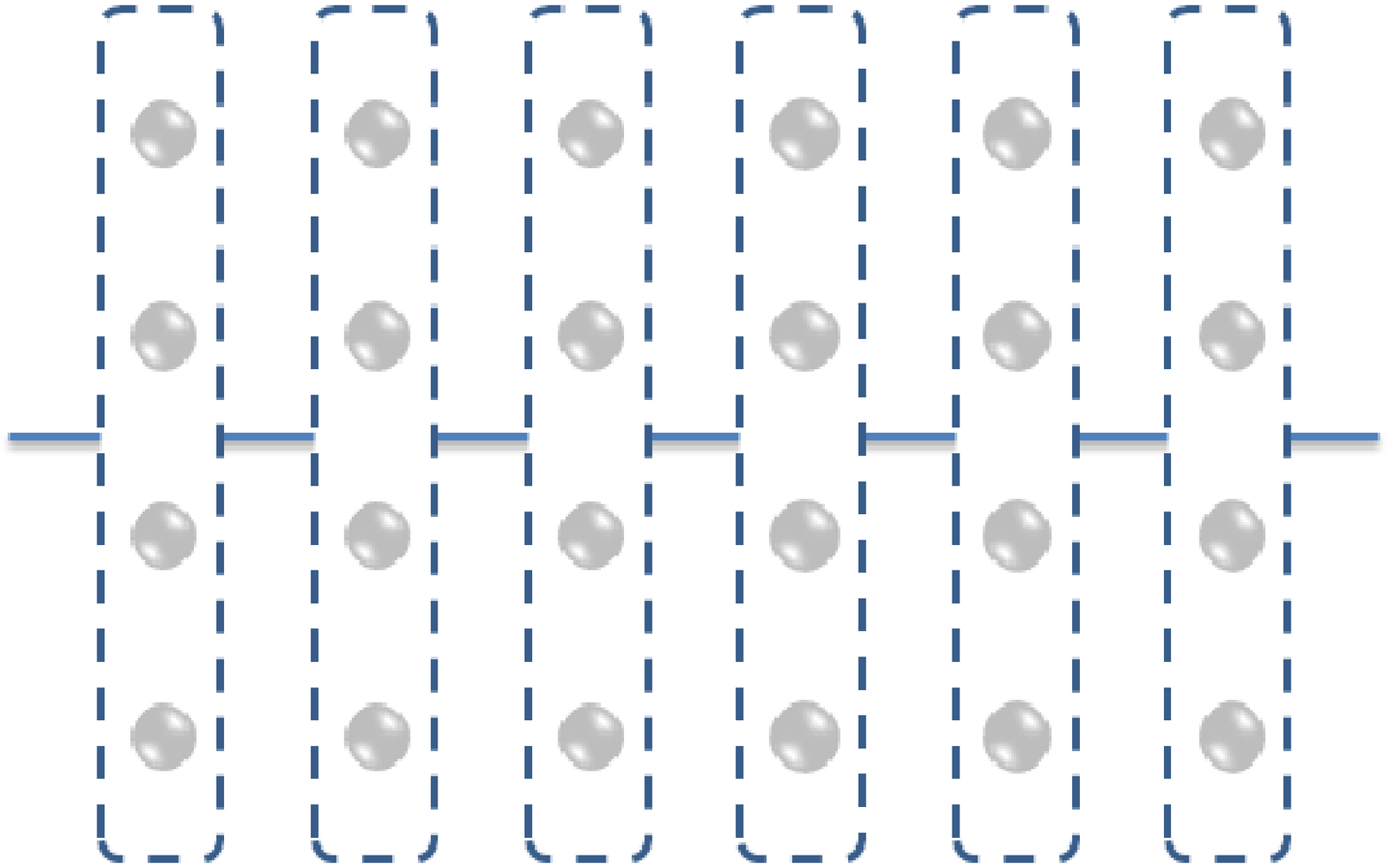}\\
		\end{array}$
		\caption{\label{fig:lattice} Spin-$\frac{1}{2}$ antiferromagnetic Heisneberg model on an infinity-by-$N$ ladder, $N=4$ for example. The circle represents lattice sites, and the lines and curves connecting the nearest neighboring sites represent the spin-spin interactions. Periodic boundary conditions are assumed. (a) Original lattice. (b) Effective lattice whose single site is indicated by the dashed rectangle enclosing the $N$ lattice sites in the rung of original lattice.}
	\end{center}
\end{figure}

\begin{figure}
	\begin{center}
		$\begin{array}{ccc}
		&\mbox{(a) TNS} &	 \mbox{(b) MPS(winding)} \\
		& \includegraphics[width=8.5pc]{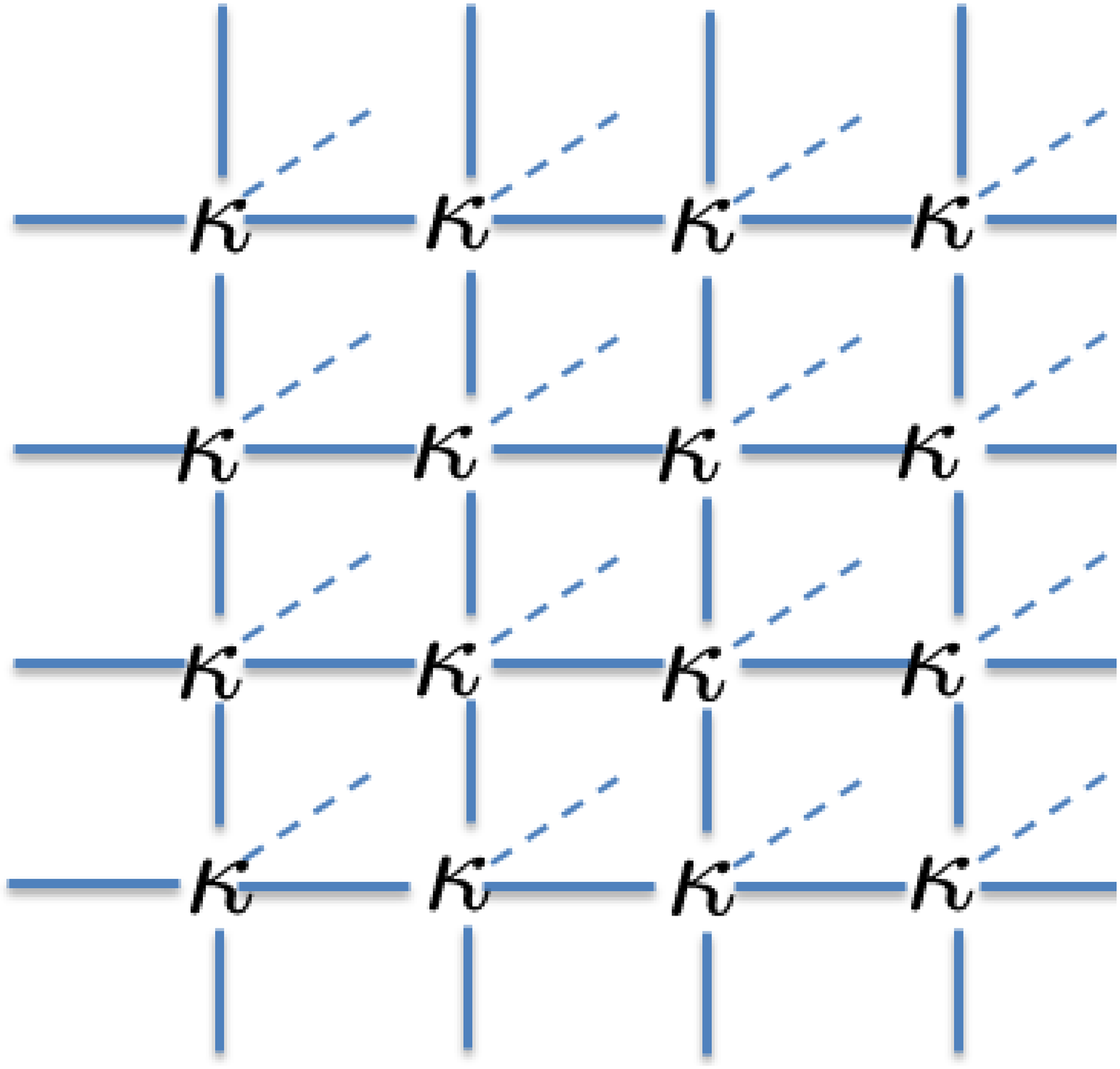} &  \includegraphics[width=8.5pc]{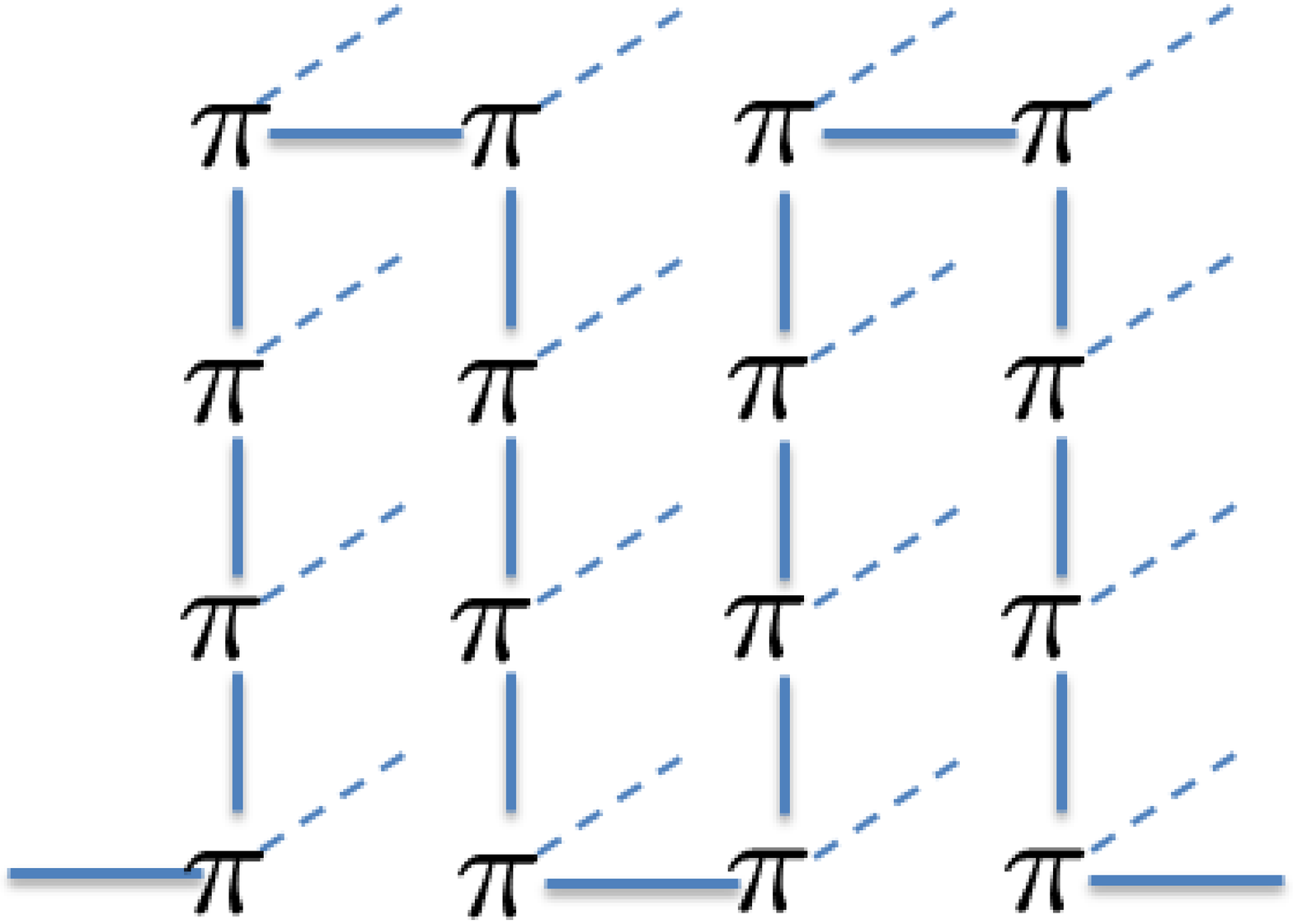}\\
		\end{array}$	
		$\begin{array}{ccc}			
		& \mbox{(c) MPS(effective 1D)}& 	 \\
  &\includegraphics[width=10.pc]{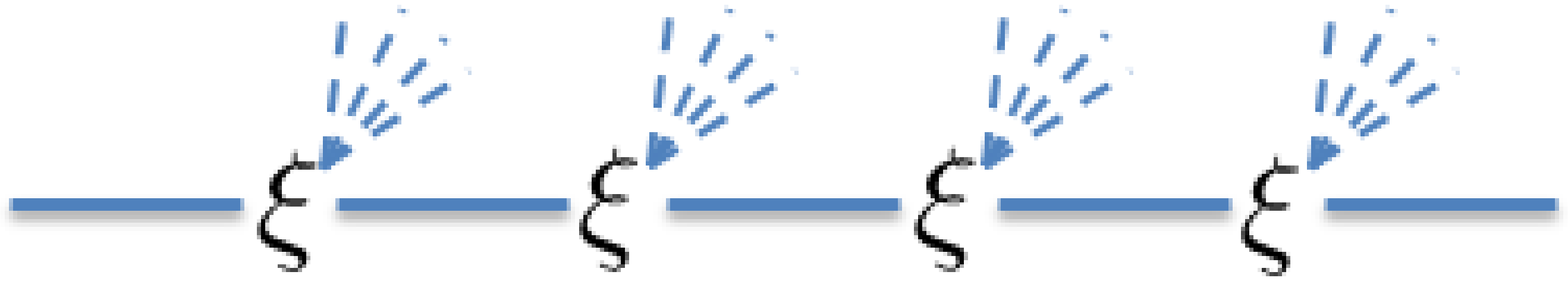}& \\
		\end{array}$
		\caption{\label{fig:TNSandMPS} Various designs for the lattice wave function using tensor/matrix product. $\xi$ denotes a tensor. The solid line refers to the bonding index while the dashed line refers to the space index. (a) Tensor network state (TNS). Each tensor has four bonding indices which resemble the same lattice architecture shown in Fig.\ref{fig:lattice} and one space index which accounts for a lattice site. TNS is employed in iPEPs, etc. (b) Matrix product state (MPS) built on a winding lattice chain. Each tensor has two bonding indices that differ from the lattice linking architecture and one space index for a site. DMRG wave function after projections is reduced to this form of MPS. (c) MPS built on a lattice chain of a translational symmetry. The two bonding indices resemble the architecture of an effective lattice shown in Fig.\ref{fig:lattice}(b); the combination of dashed lines each of which corresponds to a physical site of each $\xi$ is treated as a single space index running from 1 to $2^N$ for spin-$\frac{1}{2}$.}
	\end{center}
\end{figure}

Nevertheless, understanding the dimension transition, from the non-ordered 1D including quasi-1D spin lattices to the ordered 2D lattice, is important in taking the right numerical strategy to deal with strong correlation in low-dimensional quantum system. For instance, DMRG works extremely well in 1D but not in 2D. When dealing with a 2D lattice, the wave function obtained in DMRG has a matrix product state (MPS)\cite{Fannes1992,Oestlund1995,Verstraete2004,Chung2006,Chung2007,Chung2009,Wang2012,Garcia2007,Crosswhite2008,McCulloch2008a} form that is built on a winded 1D lattice which resembles the 2D lattice\cite{Stoudenmire2012,Ehlers2017}. See Fig.\ref{fig:TNSandMPS}(b) for an illustration. According to the area law\cite{Eisert2010}, the required MPS rank (bond index size) characterizing the entanglement in the wave function increases too rapidly in this way\cite{Schollwoeck2005}. It is obvious that rather than treating the infinity-by-$N$ lattice as a winded 1D lattice, it can also be treated as a 1D lattice by converting $N$ physical sites in the rung into an effective site\cite{Wang2015} (Fig.\ref{fig:lattice}(b)).
 We take the latter approach in this study and use the infinity-by-$N$ lattice to investigate if the system wave function, no matter which dimensional characteristics its GS turns out to be, can be universally represented by the MPS whose bonding topology is the same as the lattice linking architecture (compare the effective lattice structure in Fig.\ref{fig:lattice}(b) and the MPS structure in Fig.\ref{fig:TNSandMPS}(c)). Hence, we hopefully tame the increase of MPS rank with $N$ at a manageable rate. 

We study the model described by
\begin{equation}
\label{eq:hamiltonian}
H=J\sum_{\langle \left(i,j\right),\left(i',j'\right)\rangle}{{\vec S}_{\left(i,j\right)}\cdot {\vec S}_{\left(i',j'\right)}}. 
\end{equation} 
where ${\vec S}_{\left(i,j\right)}$ is the spin vector operator on the ${\left(i,j\right)}^{\text{th}}$  lattice site with $\it{i}$  running from $-\infty$ to $\infty$ in the longitudinal direction (LD) and $\it{j}$ running from $1$ to $N$ in the rung. $\langle\rangle$ sums over the nearest neighboring sites. $J$ is the spin-spin coupling integral and is normalized to $1$ hereafter. The periodic boundary condition (PBC) is assumed in both directions. See Fig.\ref{fig:lattice}(a) for the schematic of the lattice geometry and interaction configurations.

\subsection{\label{subsec:methodology}Methodology}
As mentioned, we divide an infinity-by-$N$ square lattice into an infinite chain of effective sites, each of which is converted from the N sites in the rung. See Fig.\ref{fig:lattice}(b) for illustration. Each effective site has $2^N$ degree of freedom. The wave function is written in a matrix product state that is built on the effective sites as,
\begin{equation}
\label{eq:mps}
\mid \psi\rangle=\sum_{\cdots r^{i-1}r^i\cdots}{tr\left(\cdots \xi_{r^{i-1}}\cdot \xi_{r^i}\cdots\right)\cdots\mid \phi_{r^{i-1}}^{i-1}\rangle\mid \phi_{r^{i}}^{i}\rangle\cdots} 
\end{equation} 
. This is composed of only one tensor $\xi$ after implementation of the antiferromagnetic checker board transformation (Sec.\ref{sec:checkerboard}). $\xi$ has three indices, as on Fig.\ref{fig:TNSandMPS}(c). The first index is associated with the local quantum state of $i^{\text{th}}$ effective site, $\mid \phi_{r^i}^i \rangle$. Therefore, it runs from $1$ to $2^N$. The other two legs are the left/right bond indices contracting with the right/left bond indices of the front/rear tensors. These two legs run from $1$ to $P$; $P$ is a chosen parameter characterizing the entanglement in the MPS wave function, the larger of which gives the more precise representation of the wave function\cite{Schollwoeck2005}.
 
Meanwhile, the Hamiltonian $H$ is transformed to a matrix product operator (MPO)\cite{Crosswhite2008,Chung2007,Chung2009,Wang2012,Wang2015,Verstraete2004a,Pirvu2010,Schollwoeck2005} via the density operator $e^{-\beta H}$ with $\beta$ being a small positive constant\cite{Chung2006,Chung2007,Wang2012,Wang2015} as
\begin{widetext}
	\begin{align}
	\label{eq:mpo}
e^{-\beta H}=&\sum_{\substack{\cdots r^{i-1}r^i\cdots\\\cdots s^{i-1}s^i\cdots}}{tr\left(\cdots \Gamma_{r^{i-1}s^{i-1},mn}\left(\beta\right)\cdot \Gamma_{r^is^i,no}\left(\beta\right)\cdots\right) \cdots\mid \phi_{r^{i-1}}^{i-1}\rangle\mid \phi_{r^{i}}^{i}\rangle\cdots\langle \phi_{s^{i-1}}^{i-1}\mid\langle \phi_{s^{i}}^{i}\mid\cdots}	
	\end{align}
\end{widetext}
. See Sec.\ref{sec:mpo} for details. Note that the checker board symmetry is applied as well to retain only one tensor $\Gamma\left(\beta\right)$ which has four legs. The first two legs are similar to the first leg of MPS tensor, associated with the local quantum state of $i^{\text{th}}$ effective site $\mid \phi_{r^i}^i \rangle$ and its conjugate $\langle \phi_{s^i}^i \mid$, running from 1 to $2^N$. The other two legs are the left/right bond indices. But different from MPS' bond indices, they run from 1 to $Q\equiv4^N$ (Fig.\ref{fig:mpo}, Sec.\ref{sec:mpo}). Namely, they are explicitly determined by the lattice topology and the interaction configuration.  
Both MPS and MPO are entangled quantities in that they cannot be simply expressed as a product of multiplicative terms each of which only involves local quanta. The entanglement in MPS is extensively studied\cite{Eisert2010,Schollwoeck2005}, while that of MPO is rarely addressed. Another entangled quantity, whose entanglement is either rarely addressed, is the observation of energy (via the density operator) that arises from a straightforward derivation from MPS and MPO as
\begin{widetext}
	\begin{align}
	\label{eq:observation}
	\langle \psi \mid e^{-\beta H} \mid \psi\rangle
	=&tr\left[\cdots\left(\xi_{s^{i-1},\alpha_1\delta_1}\Gamma_{s^{i-1}r^{i-1},\alpha_2\delta_2}\xi_{r^{i-1},\alpha_3\delta_3}\right)\cdot
	\left(\xi_{s^{i},\delta_1\eta_1}\Gamma_{s^{i}r^{i},\delta_2\eta_2}\xi_{r^{i},\delta_3\eta_3}\right)\cdots\right]
	\end{align}
\end{widetext}
, where each parenthesized term in the trace separated by the product is a new tensor, the bond indices between which can be combined into new compound indices that give the matrix rank in equation (\ref{eq:observation}). The matrix ranks in equations (\ref{eq:mps}), (\ref{eq:mpo}), and (\ref{eq:observation}) characterize the entanglement in MPO, MPS and energy observation, respectively. It is clear that the rank of the last quantity is determined by the rank of the MPS and that of the MPO together. Explicitly, it equals to $QP^2$. The entanglement in $\langle e^{-\beta H}\rangle$ is important for controlling the burden of numerical simulation because this observation is a precursor step to variationally optimizing the wave function (See Sec.\ref{sec:mps}). It leads to a singular value decomposition (SVD) of rank of $QP^2$ for the building unit of $\langle e^{-\beta H}\rangle$. The fast increase of the rank with $N$ dominates the other processes, as explained below.

On the other hand, varying $\langle e^{-\beta H}\rangle$ with respect to the MPS tensor $\xi$ yields a generalized eigenvalue equation (GEE) of rank of $2^NP^2$, where $2^N$ accounts for the local quantum space, as will be explained in Sec.\ref{sec:mps}. GEE is formed using the trial MPS wave function at the very beginning and then is updated by the solved eigenvector corresponding to the largest eigenvalue at each iteration. Eventually the eigenvector approaches the fixed state for a given rank $P$. Adding small new elements to the obtained MPS matrix ranked at $P$ to form a new trial MPS matrix ranked slightly larger at $P+\Delta P$, we carry on the previous process till convergence. Thus, those obtained quantities will converge with $P$. The final largest eigenvalue, i.e., $e^{-\beta \epsilon_0}$, gives the GS energy $\epsilon_0$. 

Obviously, for $N>1$, the rank $QP^2$ of SVD dominates over the rank $2^NP^2$ of GEE. Nevertheless, we show that the essential concept that distinguishes iqEPT from other MPS methods is that it expresses the Hamiltonian as a parameterized MPO\cite{Chung2006,Chung2007,Wang2012} and this parameter was used to reduce the linking complexity between the building units of MPO\cite{Wang2015}. In this work, we further point out that it is equivalent to treat the entanglement in MPO in perturbation, as will be explained in Sec.\ref{sec:ept}. In fact, owing to the small positive parameter $\beta$,
\begin{equation}
\label{eq:ompept}
\Gamma\left(\beta\right)=\Gamma_0\left(\beta\right)+\Gamma_2\left(\beta^2\right)
\end{equation}
where the first term on the right-hand side collects the elements in MPO that is in the zeroth and first orders of $\beta$, while the second term includes terms of higher orders. If one sets $\beta$ to be a value as small as $10^{-7}$, only $\Gamma_0$ needs to be retained. A merit is that $\Gamma_0$ is extremely sparse. Hence, it can be reduced to a much smaller tensor whose rank (bond index size) is $3N+1$, in contrast to the original scaling of $4^N$, see Sec.\ref{sec:ept}. After this major leap to the reduction of MPO rank $Q$, GEE's scale $2^NP^2$ becomes dominant over the new rank $\left(3N+1\right)P^2$ of SVD. Then, we integrate the Jacobi-Davidson method for GEE\cite{Sleijpen1996} with both MPO and MPS, without explicitly forming GEE. The details are given in Sec.\ref{sec:davidson}. As a result, we were able to handle the unprecedented GEE rank as large as $2^{14} \times 350^2 = 2.0 \times 10^9$ when $P = 350$ for $N = 14$.

We organize the remainder of discussion as follows. We first discuss the parameterized MPO for an infinity-by-$N$ spin-$\frac{1}{2}$ antiferromagnetic Heisenberg model in Sec.\ref{sec:mpo} and discuss the variation of MPS in the presence of MPO in Sec.\ref{sec:mps}. Next, we discuss the entanglement perturbation of MPO and of Hilbert space in Sec.\ref{sec:ept}. There, we show the area law is not applicable in this study. Implementation of the antiferromagnetic checkerboard symmetry is given in Sec.\ref{sec:checkerboard} to simplify the formulations in the previous sections. It is followed by the introduction of integration of the Davidson eigenvalue solver with MPS and MPO in Sec.\ref{sec:davidson}. A useful relationship between the spin-spin correlations and the staggered magnetization is discussed in Sec.\ref{sec:correlation}. There, we also demonstrate that $Ln\left(LnC_r-LnC_{r+1}\right)$, $C_r$ being the spin-spin correlation at separation $r$, can be used to interpret order/disorder. Sec.\ref{sec:reduction} introduces the space reduction in MPS. Sec.\ref{sec:result} discusses the results. Conclusion is reached in Sec.\ref{sec:Conclusion} with an outlook in Sec.\ref{sec:outlook}.

\section{\label{sec:mpo}Parameterized Matrix product operator for Heisenberg spin-$\frac{1}{2}$ model on infinite by N square lattices}
\begin{figure}
	\begin{center}
		$\begin{array}{cc}
		&\mbox{(a) construction of MPO}  \\
		& \includegraphics[width=20.pc]{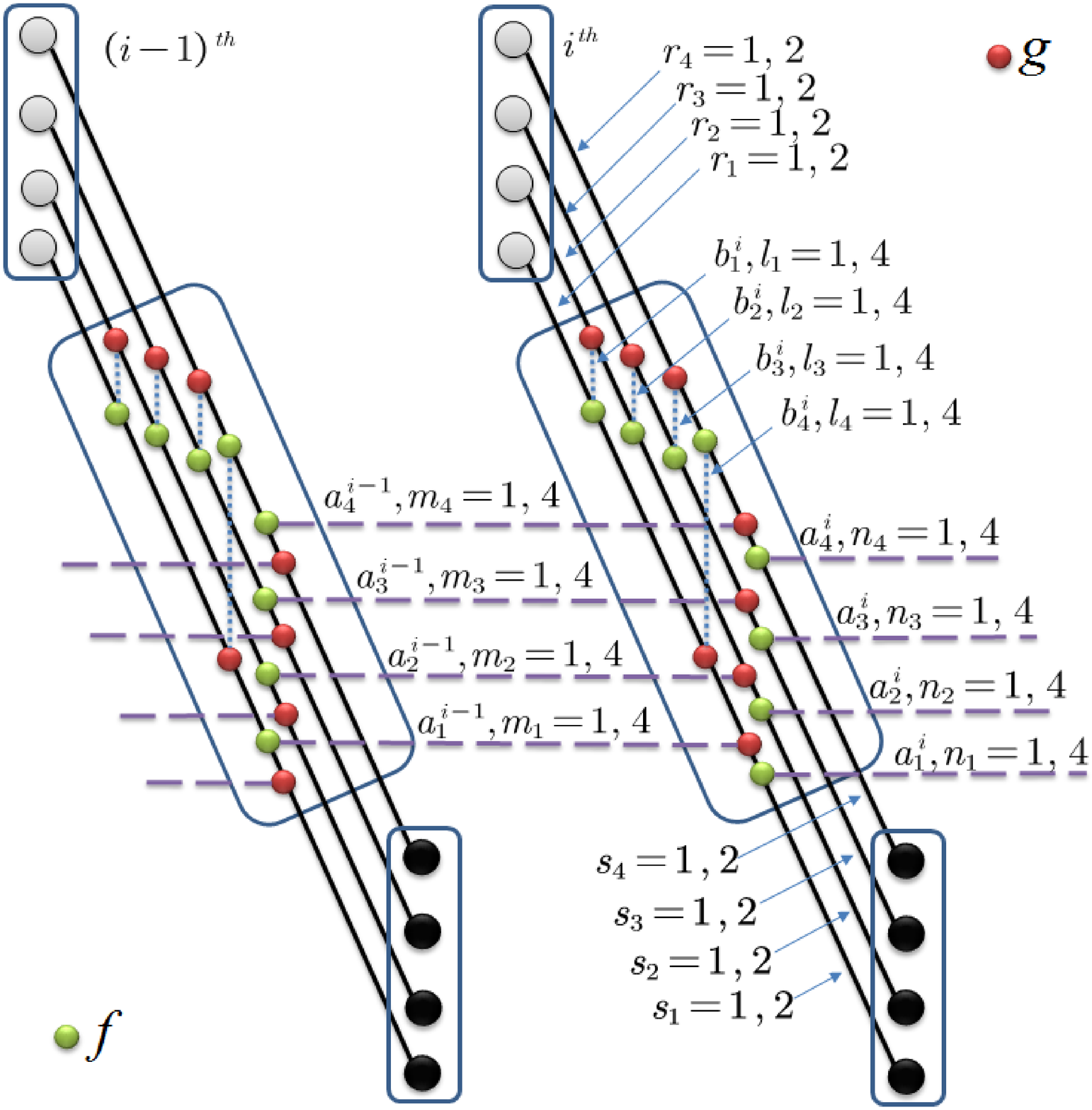}\\
		& \mbox{(b) symbolic MPO}\\
		& \includegraphics[width=13.pc]{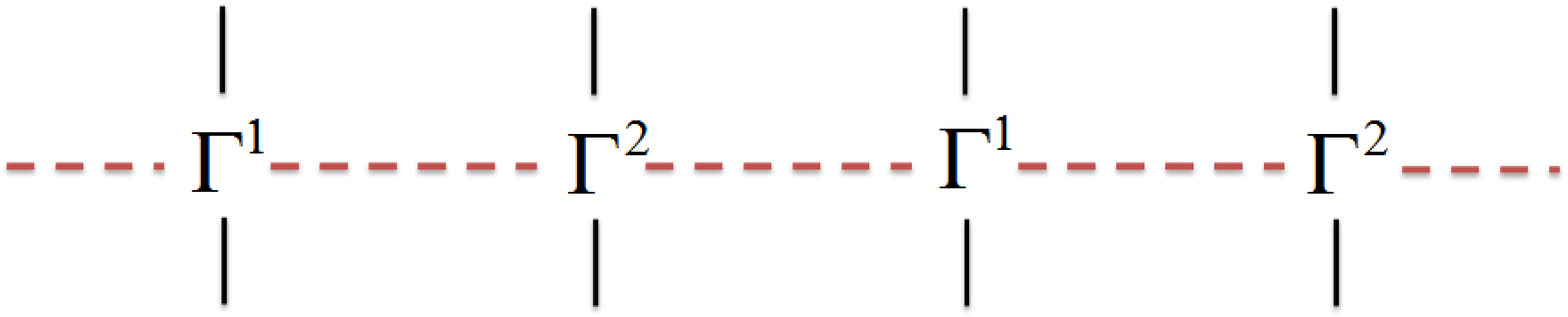} 
		\end{array}$
		\caption{\label{fig:mpo} Illustration of MPO for $N=4$. (a) The construction. The local quantum space is represented by rectangles enclosing $4$ physical sites (circle in gray color). Out of page is the conjugate of those spaces. The building units of MPO , symbolized by $\Gamma^1$ and $\Gamma^2$ in (b), are enclosed within the tilted rectangles in between the spaces. Going from the space to its conjugate, the first four pairs of green-red solid circles are stacked shell-by-shell in sequence. They account for the on-effective-site interaction, with the green/red circle denoting $f$/$g$ operator mentioned in the context. The dashed line denotes the contraction of bond index running from $1$ to $4$. The solid line denotes the inner product between the operators operating sequentially on the same physical lattice site. Following the on-effective-site operators are one shell of $f$/$g$ operators bonding the $\left(i-1\right)^{th}$/$i^{th}$ effective sites, and another shell of $g$/$f$ operators bonding the $\left(i-2\right)^{th}$/$\left(i-1\right)^{th}$ effective sites and the $i^{th}$/$\left(i+1\right)^{th}$ effective sites, respectively. They are stacked layer-by-layer from bottom to top, forming inter-effective-site interactions and hence giving the rise of entanglement in MPO represented schematically by the horizontal dashed lines in (b). The individual space indices $r_{i=1,4}$/$s_{i=1,4}$ are combined into single indices of vertical solid lines in (b); while the individual bond indices $m_{i=1,4}$/$n_{i=1,4}$ being combined into horizontal dashed lines in (b).}
	\end{center}
\end{figure} 
In what follows, the Einstein summation convention is implied for repeating indices in a formula except stated otherwise. $\left(x\leftrightarrow y\right)$ denotes pairing between two physical sites $x$ and $y$. $\lfloor \alpha_1, \cdots ,\alpha_m\rfloor$ combines $m$ individual indices $\alpha_1=1,\cdots k_1,\cdots , \alpha_m=1,\cdots, k_m$ into a single flattened index $\alpha=1,\cdots ,\prod_{i}^m{k_i}$.

After the physical sites in the rung are converted into single effective sites, the Hamiltonian is further rewritten in the following form,
\begin{equation}
\label{eq:hbond1}
H=\sum_{\{a,b\}}{\left(H_a+H_b\right)}
\end{equation}
, where the summation runs over the two sets of bonds, $\{a\}$ and $\{b\}$, between nearest neighboring physical sites. If the two physical sites of a bond reside on different effective sites, it is collected in the inter-effective-site set $\{a\}$; otherwise in the intra-effective-site set $\{b\}$. Fig.\ref{fig:mpo} illustrates various bonds in the case of $N=4$. In this case, each effective site, say the $i^{th}$ site, has four intra-effective-site bonds $b_1^i$, $b_2^i$, $b_3^i$ and $b_4^i$. It also participates in eight inter-effective-site bonds. The first four are labeled with index $i-1$: $a_1^{i-1}$, $a_2^{i-1}$, $a_3^{i-1}$ and $a_4^{i-1}$. They bond physical sites residing on the $\left(i-1\right)^{th}$ and $i^{th}$ effective sites. The last four are labeled with index $i$; they are $a_1^i$, $a_2^i$, $a_3^i$ and $a_4^i$ bonding physical sites residing on the $i^{th}$ and $\left(i+1\right)^{th}$ effective sites. These sets of bonds are used to rewrite the Hamiltonian as follows,
\begin{align}
\label{eq:density}
e^{-\beta H} \approx \prod_{i}{\left(\prod_{k}{e^{-\beta H_{a_k^i}}}\prod_{l}{e^{-\beta H_{b_l^i}}}\right)}+\bigcirc\left(\beta^{2}\right)
\end{align}
where $i=1,\cdots ,\infty$ and $k,l=1,\cdots ,N$. Although the sequence of the bonds grouped as the single exponent in the left hand side of equation (\ref{eq:hbond1}) does not matter, it matters in the right hand side in that the ordering of $i$, $k$, $l$ is equivalent to permuting the Hamiltonian matrix. The permutation does not affect the physical property but requires the corresponding linear manipulation of the representation basis. We choose to operate $\prod_{k}{e^{-\beta H_{a_k^i}}}$ on $\prod_{l}{e^{-\beta H_{b_l^i}}}$ for given $i$. A physical site of the $i^{th}$ effective site is denoted as $x^i$, $x=1,\cdots ,N$ from bottom to top in the rung. For the set $\{b^i_l\}$, it is ordered such that $l=1:\left(1^i \leftrightarrow 2^i\right)$, $2:\left(2^i \leftrightarrow 3^i\right),\cdots,$ $N:\left(N^i \leftrightarrow 1^i\right)$. For the set $\{a^i_k\}$, it is ordered such that $k=1:\left(1^i \leftrightarrow 1^{i+1}\right)$, $2:\left(2^i \leftrightarrow 2^{i+1}\right),\cdots,$ $N:\left(N^i \leftrightarrow N^{i+1}\right)$. It is clear that the successive product of $e^{-\beta H_{b_l^i}}$ involves only one effective site. The $N$ operations are stacked in a shell-by-shell manner in and out of the page. It is followed by the successive product of $e^{-\beta H_{a_k^i}}$ which operates on the physical sites across two effective sites. They are stacked from bottom to top in a layer-by-layer manner. See Fig.\ref{fig:mpo} for details. Each individual density operator for a general bond $\left(i\leftrightarrow j\right)$ in the right side of equation (\ref{eq:density}) is Taylor-expanded, utilizing small positive $\beta$,
\begin{align}
\label{eq:density2}
e^{-\beta\left(S^x_i S^x_j+S^y_i S^y_j+S^z_i S^z_j\right)}\approx &I-\beta\left(S^x_i S^x_j+S^y_i S^y_j+S^z_i S^z_j\right)\notag\\
\equiv &f_{\alpha}\otimes g_{\alpha}
\end{align}
. $f_{\alpha=1,4}$ operating on the first physical site of a bond are 2 x 2 matrices (Identity, $\sqrt{\beta}S^x$, $\sqrt{\beta}{\bar{S}}^y$ and $\sqrt{\beta}S^z$); $g_{\alpha=1,4}$ operating on the second physical site are 2 x 2 matrices (Identity, $-\sqrt{\beta}S^x$, $\sqrt{\beta}{\bar{S}}^y$ and  $-\sqrt{\beta}S^z$). Note, $g$'s differ from $f$'s in that the former has the minus sign; ${\bar{S}}^y_i\otimes{\bar{S}}^y_j=S^y_i\otimes S^y_j$ with ${\bar{S}}^y$ being a real version of $S^y$ to avoid any complex element in the matrix. 

Now consider the product of two individual density operators of two bonds $c$ and $c'$. There follows two rules applied in different cases.

Rule A. 
$e^{-\beta H_c}e^{-\beta H_{c'}}=f_{\alpha}\otimes g_{\alpha}\otimes f_{\gamma}\otimes g_{\gamma}$ if there is no shared physical site.

Rule B. 
$e^{-\beta H_c}e^{-\beta H_{c'}}=f_{\alpha}\otimes \left( g_{\alpha}\cdot f_{\gamma}\right) \otimes g_{\gamma}$ if there is a shared physical site.

Rule A is transparent. The formula shown in Rule B applies to the case where the shared site is the first site of bond $c^{\prime}$ and the second site of bond $c$. Nevertheless, the combination of the positions of the shared site respectively in $c$ and $c'$ is diverse, such as first vs first, second vs second, etc. They all appear in Fig.\ref{fig:mpo}. The formula in Rule B will be slightly adjusted accordingly. 

After applying these rules, the density operator involving the $i^{th}$ effective site is
\begin{align}
\label{eq:omp}
			\left(f_{n_1}\cdot g_{m_1}\cdot g_{l_4}\cdot f_{l_1}\right)\notag\\
	\otimes \left(f_{n_2}\cdot g_{m_2}\cdot f_{l_2}\cdot g_{l_1}\right)\notag\\
	\otimes \left(f_{n_3}\cdot g_{m_3}\cdot f_{l_3}\cdot g_{l_2}\right)\notag\\
	\otimes \left(f_{n_4}\cdot g_{m_4}\cdot f_{l_4}\cdot g_{l_3}\right)
	\end{align}
. Each of the index of bond set $\{b^i\}$, $l_1$, $l_2$, $l_3$ and $l_4$, appears twice in expression (\ref{eq:omp}), implying self-contraction of intra-effective-bonds. The un-contracted indices $m_1$, $m_2$, $m_3$, and $m_4$ entangle the $i^{th}$ effective site with the $\left(i-1\right)^{th}$ effective site through the bond set $\{a^{i-1}\}$;  The other un-contracted indices $n_1$, $n_2$, $n_3$, and $n_4$ entangle the $i^{th}$ effective site with the $\left(i+1\right)^{th}$ effective site through the bond set $\{a^i\}$. In fact, fixing indices $m's$ and $n's$ to $\{m\}_0$ and $\{n\}_0$, each resultant quantity in every parenthesis of (\ref{eq:omp}) is a local density matrix $\left(\rho_{r_u,s_u}\right)$. $u=1,\cdots,4$ refer to the four parenthesis. Each of $r's$ or $s's$ runs from $1$ to $2$ accounting for spin-$\frac{1}{2}$. The combinations $r^i\equiv\lfloor r_1,r_2,r_3,r_4\rfloor$ and $s^i\equiv \lfloor s_1,s_2,s_3,s_4\rfloor$ run from 1 to $2^N$. The direct product between the four local density matrices spans a resultant density matrix $\Gamma_{r^is^i,\{m\}_0\{n\}_0}$ of rank of $2^N$ for the $i^{th}$ effective site. Allowing $m$'s and $n$'s to vary, $\Gamma$ becomes a four-leg tensor $\Gamma_{r^is^i,mn}$. The combinations $m\equiv\lfloor m_1,m_2,m_3,m_4\rfloor$ and $n\equiv \lfloor n_1,n_2,n_3,n_4\rfloor$ are the bond indices and run from 1 to $4^N$. Considering the bipartite structure due to the antiferromagnetic nature, equation \eqref{eq:density} is transformed to a parameterized MPO as follows,
\begin{widetext}
	\begin{equation}
	\label{eq:omp2}
	e^{-\beta H}=\sum_{\substack{\cdots r^{i-1}r^i\cdots\\\cdots s^{i-1}s^i\cdots}}{tr\left(\cdots\Gamma_{r^{i-1}s^{i-1},mn}^1\left(\beta\right)\cdot\Gamma_{r^is^i,no}^2\left(\beta\right)\cdots\right)\cdots\mid \phi_{r^{i-1}}^{i-1}\rangle\mid \phi_{r^i}^i\rangle\cdots \langle \phi_{s^{i-1}}^{i-1}\mid \langle\phi_{s^i}^i\mid\cdots }
	\end{equation}
\end{widetext}
. Note that, in equations \eqref{eq:density} and \eqref{eq:omp2} the small parameter $\beta$ is used to express $e^{-\beta H}$ as a successive product of operators according to BCH formula, implying that parts of high-order terms in $\beta$ are already omitted controllably. Nevertheless, many other high-order terms in $\beta$ are still remaining in the operator, which is a consistency check of formula (\ref{eq:omp}). In what follows in Sec.\ref{sec:mps}, we show that the remaining high-order terms in $\beta$ could be treated perturbatively to reduce the last two (bond) indices of $\Gamma^{1,2}$ from $4^N$ to $3N+1$. But for the moment we first discuss how to variationally optimize the MPS wave function in the presence of MPO.

\section{\label{sec:mps}Variational optimization of MPS in the presence of MPO }
The system wave function is expressed as a MPS
\begin{equation}
\label{eq:mps1}
\mid\Psi\rangle=\sum_{\cdots r^{i-1}r^i\cdots}{tr\left(\cdots\xi^1_{r^{i-1}}\cdot\xi^2_{r^i}\cdots\right)\cdots\mid\phi_{r^{i-1}}^{i-1}\rangle\mid\phi_{r^i}^i\rangle\cdots}
\end{equation}
, which has the $\xi^{1,2}$ repetition structure, similar to that for MPO, due to the antiferromagnetic condition. Note that $\xi^{1,2}$ are 3-leg tensors. For example, $\xi^2_{r^i}$ for the $i^{th}$ effective site is explicitly denoted with the index of the first leg, $r^i=1,2,\cdots ,2^N$, which refers to the local quantum state $\mid \phi_{r^{i}}^i\rangle$. Explicitly writing down the other two legs, $m$ and $n$ in $\xi_{r^i,mn}^2$ are the left/right indices of a matrix, fixing $r^i$. Therefore, each of $\xi^{1,2}$ has $2^NP^2$ variables when the matrix rank is $P$. Given a configuration (Fock vector) of the local state of all effective sites, a specific $P \times P$ matrix is then assigned to each effective site; the left/right indices of that matrix contract in a closed form with the right/left indices of the matrix of the front/rear effective sites to yield a trace-out. The resultant scalar value is the superposition coefficient of the configuration in the wave function. Optimization of the wave function by pinpointing the superposition coefficient is equivalent to optimization of those $P\times P$ matrices, $2\times 2^N$ in total, in the MPS with the bipartition structure. 

\begin{figure}
	\begin{center}
		$\begin{array}{ccc}
		&\mbox{(a) energy observation} & \mbox{(b) normalization} \\
		& \includegraphics[width=9.5pc]{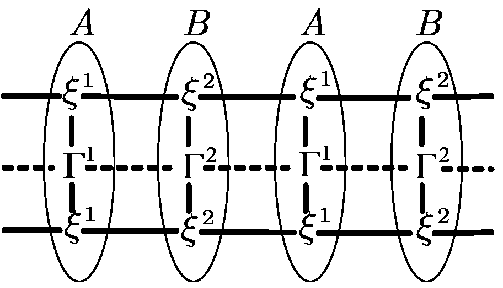}& \includegraphics[width=9.5pc]{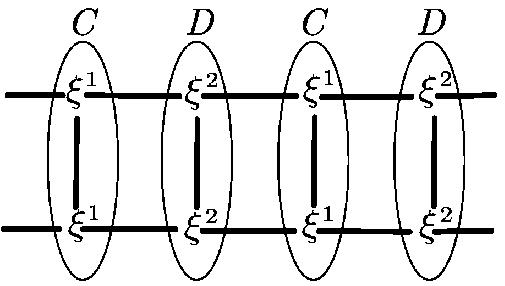}\\
		\end{array}$
		\caption{\label{fig:observation} (a) Energy observation in the presence of MPS and MPO. The quantities enclosed in an eclipse is combined into new tensors A/B, each of which has a larger rank of $P^2Q$ with $P$ and $Q$ being the MPS and MPO ranks, respectively. (b) Normalization of MPS wave function. The quantities enclosed in an eclipse is combined into new tensors C/D, each of which has a rank of $P^2$.}
	\end{center}
\end{figure}

The MPS matrices can be optimized in various ways. One way used in DMRG is to start with an exact solution of a small part of system and then to renormalize the representation basis every time the new parts interacting with the processed part are added. The MPS matrix is a fixed point after many projections of the DMRG solution\cite{Schollwoeck2005}. The other way is to variate the energy observation with respect to the MPS matrices, hence to optimize them simultaneously. Illustrated in Fig.\ref{fig:observation}, the energy observation is expressed as
\begin{widetext}
	\begin{align}
	\label{eq:observation2}
	&\langle \psi \mid e^{-\beta H} \mid \psi\rangle\notag\\
	=&\left[\sum_{\cdots s^{i-1}s^i\cdots}{tr\left(\cdots\xi^1_{s^{i-1},\alpha_1\gamma_1}\cdot\xi^2_{s^i,\gamma_1\eta_1}\cdots\right)\cdots\langle\phi_{s^{i-1}}^{i-1}\mid\langle\phi_{s^i}^i\mid\cdots}\right]\left[\sum_{\cdots r^{i-1}r^i\cdots}{tr\left(\cdots\xi^1_{r^{i-1},\alpha_3\gamma_3}\xi^2_{r^i,\gamma_3\eta_3}\cdots\right)\cdots\mid\phi_{r^{i-1}}^{i-1}\rangle\mid\phi_{r^i}^i\rangle\cdots}\right]\notag\\
	& \left[\sum_{\substack{\cdots z^{i-1}z^i\cdots\\\cdots w^{i-1}w^i\cdots}}{tr\left(\cdots\Gamma_{z^{i-1}w^{i-1},\alpha_2\gamma_2}^1\Gamma_{z^iw^i,\gamma_2\eta_2}^2\cdots\right)\cdots\mid \phi_{z^{i-1}}^{i-1}\rangle\mid \phi_{z^i}^i\rangle\cdots \langle \phi_{w^{i-1}}^{i-1}\mid \langle\phi_{w^i}^i\mid\cdots }\right]\notag\\
=&tr\left(\cdots A_{\alpha\gamma}B_{\gamma\eta}\cdots\right)
	\end{align}
\end{widetext}
where the Kronecker delta function $\langle \phi_{s^i}^i \mid \phi_{z^i}^i \rangle=\delta_{s^i,z^i}$, etc, is used to reduce the summations. And,
\begin{align}
\label{eq:relabel}
A_{\alpha\equiv\lfloor\alpha_1\alpha_2\alpha_3\rfloor ,\gamma\equiv\lfloor\gamma_1\gamma_2\gamma_3\rfloor}\equiv &\xi_{s^{i-1},\alpha_1\gamma_1}^1\Gamma_{s^{i-1}r^{i-1},\alpha_2\gamma_2}^1\xi^1_{r^{i-1},\alpha_3\gamma_3}\notag\\
B_{\gamma\equiv\lfloor\gamma_1\gamma_2\gamma_3\rfloor ,\eta\equiv\lfloor\eta_1\eta_2\eta_3\rfloor}\equiv &\xi_{s^i,\gamma_1\eta_1}^2\Gamma_{s^ir^i,\gamma_2\eta_2}^2\xi_{r^i,\gamma_3\eta_3}^2
\end{align}
. $\alpha\equiv\lfloor\alpha_1\alpha_2\alpha_3\rfloor$ denotes the combination of indices $\alpha_1,\alpha_3=1,2,\cdots,P$ and $\alpha_2=1,2,\cdots,4^N$, giving rise to a single index  $\alpha=1,2,\cdots,4^N P^2$, etc. Meanwhile, the normalization factor is
\begin{widetext}
	\begin{align}
	\label{eq:normalization}
	&\langle \psi \mid  \psi\rangle\notag\\
	=&\left[\sum_{\cdots s^{i-1}s^i\cdots}{tr\left(\cdots\xi^1_{s^{i-1},\alpha_1\gamma_1}\xi^2_{s^i,\gamma_1\eta_1}\cdots\right)\cdots\langle\phi_{s^{i-1}}^{i-1}\mid\langle\phi_{s^i}^i\mid\cdots}\right]\left[\sum_{\cdots r^{i-1}r^i\cdots}{tr\left(\cdots\xi^1_{r^{i-1},\alpha_3\gamma_3}\xi^2_{r^i,\gamma_3\eta_3}\cdots\right)\cdots\mid\phi_{r^{i-1}}^{i-1}\rangle\mid\phi_{r^i}^i\rangle\cdots}\right]\notag\\
	=&tr\left(\cdots C_{\alpha\gamma}D_{\gamma\eta}\cdots\right)
	\end{align}
\end{widetext}
. Matrices C and D are formed as
\begin{align}
\label{eq:relabel1}
C_{\alpha\equiv\lfloor\alpha_1\alpha_3\rfloor ,\gamma\equiv\lfloor\gamma_1\gamma_3\rfloor}\equiv &\xi_{r^{i-1},\alpha_1\gamma_1}^1\xi^1_{r^{i-1},\alpha_3\gamma_3}\notag\\
D_{\gamma\equiv\lfloor\gamma_1\gamma_3\rfloor ,\eta\equiv\lfloor\eta_1\eta_3\rfloor}\equiv &\xi_{r^i,\gamma_1\eta_1}^2\xi_{r^i,\gamma_3\eta_3}^2
\end{align}
where the combination of indices $\alpha_1,\alpha_3=1,2,\cdots,P$ gives rise to a single index $\alpha=1,2,\cdots,P^2$, etc. Equation (\ref{eq:observation2}) is rewritten as
\begin{equation}
\label{eq:observation3}
\langle e^{-\beta H}\rangle=\frac{\langle \psi \mid e^{-\beta H} \mid \psi\rangle}{\langle \psi \mid  \psi\rangle}=\lim\limits_{M\rightarrow\infty}{\frac{tr\left(AB\right)^M}{tr\left(CD\right)^M}}
\end{equation}
. Then, the first derivative of $\langle e^{-\beta H}\rangle$ with respect to $\xi^{1,2}$, say $\xi^1$, leads to
\begin{equation}
\label{eq:derivative}
\frac{\partial \langle \psi \mid e^{-\beta H} \mid \psi\rangle}{\partial \xi^1}=\langle e^{-\beta H}\rangle\frac{\partial \langle \psi \mid  \psi\rangle}{\partial \xi^1}
\end{equation}
Substituting the numerator and denominator in the right hand side of equation (\ref{eq:observation3}) into both sides of equation (\ref{eq:derivative}) respectively, we arrive at 
\begin{equation}
\label{eq:derivative1}
tr\left[\left(AB\right)^{M-1}\frac{\partial A}{\partial \xi^1}B\right]=\langle e^{-\beta H}\rangle tr\left[\left(CD\right)^{M-1}\frac{\partial C}{\partial \xi^1}D\right]
\end{equation}
. Singular-value-decomposing the building units in equation (\ref{eq:observation3}), we have
\begin{align}
\label{eq:ABCD}
AB=& u\Lambda v\notag\\
CD=& u'\Delta v'
\end{align}
Substituting equation (\ref{eq:ABCD}) into equation (\ref{eq:derivative1}) derives that
\begin{equation}
\label{eq:derivative2}
v\frac{\partial A}{\partial \xi^1}Bu=\langle e^{-\beta H}\rangle\left(\frac{\Delta_1}{\Lambda_1}\right)^{M-1}v'\frac{\partial C}{\partial \xi^1}Du'
\end{equation}
In the above, only the largest eigenvalues $\Lambda_1\left(\Delta_1\right)$ and the corresponding right/left eigenvectors $v/u$ ($v/u'$) survive when $M\rightarrow \infty$. On the other hand, substituting equation (\ref{eq:ABCD}) into equation (\ref{eq:observation3}) leads to 
\begin{equation}
\label{eq:derivative3}
\langle e^{-\beta H}\rangle^{\frac{1}{M}}=\frac{\Lambda_1}{\Delta_1}
\end{equation}
Substituting equation (\ref{eq:derivative3}) into equation (\ref{eq:derivative2}), it arrives at
\begin{equation}
\label{eq:derivative4}
v\frac{\partial A}{\partial \xi^1}Bu=\langle e^{-\beta H}\rangle^{\frac{1}{M}}v'\frac{\partial C}{\partial \xi^1}Du'
\end{equation}
In fact equation (\ref{eq:derivative4}) is a generalized eigenvalue equation. To confirm it, it is instructive to explicitly rewrite both sides as
\begin{align}
\label{eq:gee}
v\frac{\partial A}{\partial \xi^1}Bu=&X_{\lfloor r^{i-1},\alpha_1,\gamma_1\rfloor ,\lfloor s^{i-1},\alpha_3,\gamma_3\rfloor}\chi^1_{\lfloor s^{i-1},\alpha_3,\gamma_3\rfloor}&=X\chi^1\notag\\
v'\frac{\partial C}{\partial \xi^1}Du'=&Y_{\lfloor r^{i-1},\alpha_1,\gamma_1\rfloor ,\lfloor s^{i-1},\alpha_3,\gamma_3\rfloor}\chi^1_{\lfloor s^{i-1},\alpha_3,\gamma_3\rfloor}&=Y\chi^1
\end{align}
where $\chi^1_{\lfloor s^{i-1},\alpha_3,\gamma_3\rfloor}$ is the vector version of 3-leg tensor $\xi^1_{s^{i-1},\alpha_3\gamma_3}$. And,
\begin{align}
\label{eq:gee1}
&X_{\lfloor r^{i-1},\alpha_1,\gamma_1\rfloor ,\lfloor s^{i-1},\alpha_3,\gamma_3\rfloor}\notag\\
=& v_{\alpha_1\alpha_2\alpha_3}\Gamma_{r^{i-1}s^{i-1},\alpha_2\gamma_2}^1\xi_{r^{i},\gamma_1\eta_1}^2\Gamma^2_{r^{i}s^{i},\gamma_2\eta_2}\xi_{s^{i},\gamma_3\eta_3}^2 u_{\eta_1\eta_2\eta_3}\notag\\
&Y_{\lfloor r^{i-1},\alpha_1,\gamma_1\rfloor ,\lfloor s^{i-1},\alpha_3,\gamma_3\rfloor}\notag\\
=& {v'}_{\alpha_1\alpha_3}\delta_{r^{i-1},s^{i-1}}\xi_{r^{i},\gamma_1\eta_1}^2\xi_{r^{i},\gamma_3\eta_3}^2 {u'}_{\eta_1\eta_3}
\end{align}
are matrices. Thus,
\begin{align}
\label{eq:gee2}
X\chi^1=&\langle e^{-\beta H}\rangle^{\frac{1}{M}}Y\chi^1\notag\\
X'\chi^2=&\langle e^{-\beta H}\rangle^{\frac{1}{M}}Y'\chi^2
\end{align}
where another generalized eigenvalue equation is generated for $\chi^2$, the vector version of the 3-leg tensor $\xi^2$. It is straightforward that $X/X'$ and $Y/Y'$ are functional of $\chi^2/\chi^1$. If we start to generate the GEEs with the trial vectors $\chi_{0,P_0}^{1,2}$, the rank of each being as small as $2^NP_0^2$, a new set of vectors $\chi_{1,P_0}^{1,2}$ are obtained by solving those GEEs and are used to update the GEEs. Iterations are carried on until the norm of $\| \chi_{m+1,P_0}^{1,2}-\chi_{m,P_0}^{1,2} \|$ is less than a threshold value. The converged vectors for $P_0$ are used to generate the trial vectors for a slightly larger rank $P_0+\Delta P$ by adding small new elements to the enlarged vectors. A second type convergence with respect to $P$ should eventually bring the identical largest eigenvalue $\lambda=\frac{\Lambda_1}{\Delta_1}$ for equation (\ref{eq:gee2}). We have
\begin{equation}
\label{eq:energy}
\langle e^{-\beta H}\rangle=\lambda^M
\end{equation}
. The GS energy of an infinite by N lattice is
\begin{equation}
\label{eq:energy1}
\epsilon_0=-\beta^{-1}ln{\langle e^{-\beta H}\rangle}=-M\beta^{-1}ln{\lambda}
\end{equation}
. The GS energy per spin is
\begin{equation}
\label{eq:energy2}
\bar{\epsilon}_0=\frac{\epsilon_0}{2MN}=-\left(2N\beta\right)^{-1}ln{\lambda}
\end{equation}

\section{\label{sec:ept}quasi-1D Entanglement perturbation theory for an infinite by $N$ Heisenberg square lattice }
\subsection{\label{sec:mpoept} Entanglement Perturbation in Hamiltonian Space}
There are two kinds of eigenvalue equations to be solved in this method. The first kind is the SVDs in equation (\ref{eq:ABCD}) for the left/right eigenvectors of two asymmetric matrices. The first matrix has rank of $R_1=4^NP^2$ while the second has rank of $R^{\prime}_1=P^2$; the second kind is the GEEs in equation (\ref{eq:gee2}) of rank of $R_2=2^NP^2$. It is obvious that $R_1$ dominates $R_2$ when $N>1$. The largest simulation scale in our study is for $N=14$ and $P=350$. It corresponds to $R_1=3.3\times 10^{13}$ and demands $30$ $\it{Tbyte}$ memory to store even a single eigenvector, not mentioning that solving an eigenvalue equation requires much more memory allocation than that merely to store a single eigenvector. This data scale is apparently not practical for modern computers. 

Fortunately, there is a simple way to overcome this difficulty. Examining $f_{\alpha=1,\cdots,4}$ and $g_{\alpha=1,\cdots,4}$, the $\alpha=1$ terms are the identity matrix, zeroth order of $\beta$; the terms of $\alpha=2,3,4$ are all in the order of $\sqrt{\beta}$. But whenever there is a term of order $\sqrt{\beta}$, there should be a counterpart in the same order at the other end of a bond. They actually generate terms in the first order of $\beta$. On the other hand, according to formula (\ref{eq:omp}), the bond index of $\Gamma^{1,2}$, say, $m\equiv\lfloor m_1,\cdots,m_N\rfloor$ is a combination of $m_1,\cdots,m_N$, each of which is the index of $f$ or $g$ running from $1$ to $4$. Because $\beta\le 10^{-7}$, it is safe to discard the terms in the order of $\beta^2$ and the beyond. Therefore, it amounts to keeping, among $4^N$ combinations of $\lfloor m_1,\cdots,m_N\rfloor$, those terms with at most one of $m's$ not being 1. The new bond index of $\Gamma^{1,2}$, after reduction, now runs from 1 to $3N+1$ instead from $1$ to $4^N$. In the case of $N=4$, they are $1$ for $1111$, $2$ for $1112$, $3$ for $1113$, $4$ for $1114$, $5$ for $1121$, $6$ for $1131$, etc.
  
At this point, we reduced the rank of MPO tensor in perturbation of the small positive parameter $\beta$. Now we show that the reduction of rank is corresponding to the entanglement reduction in MPO. The MPO is an entangled quantity in that it cannot be expressed as the product of individual multiplicative quantities that is associated with local quanta, namely the local state of the effective site in this study. 

Like the Hilbert space for the wave function of a quantum lattice, $\cdots\otimes\{\mid \phi_i\rangle\}\otimes\{\mid \phi_{i+1}\rangle\}\otimes\cdots$, we define a space $\mathcal{H}$, $\cdots\otimes\{\mid \phi_i\rangle\langle\phi_i^{\prime}\mid\}\otimes\{\mid \phi_{i+1}\rangle\langle\phi_{i+1}^{\prime}\mid\}\otimes\cdots$ for the Hamiltonian. It is not hard to show that it is indeed a vector space after defining an inner product, i.e., vector-vector multiplication as
\begin{align}
\label{eq:hamiltonianspace}
&\vec{h}_1\cdot\vec{h}_2\notag\\=&\cdots\left(\langle \varphi_i^{\prime}\mid\phi_i\rangle \langle \phi_i^{\prime}\mid\varphi_i\rangle\right)\left(\langle \varphi_{i+1}^{\prime}\mid\phi_{i+1}\rangle \langle 
\phi_{i+1}^{\prime}\mid\varphi_{i+1}\rangle\right)\cdots\\
&\begin{array}{c}
\forall\vec{h}_1,\vec{h}_2\in \mathcal{H}\\
\vec{h}_1=\cdots\otimes\{\mid \phi_i\rangle\langle\phi_i^{\prime}\mid\}\otimes\{\mid \phi_{i+1}\rangle\langle\phi_{i+1}^{\prime}\mid\}\otimes\cdots\notag\\
\vec{h}_2=\cdots\otimes\{\mid \varphi_i\rangle\langle\varphi_i^{\prime}\mid\}\otimes\{\mid \varphi_{i+1}\rangle\langle\varphi_{i+1}^{\prime}\mid\}\otimes\cdots
\end{array}
\end{align}
. A MPO is exactly a vector in the form of MPS in $\mathcal{H}$. Analogously, the entanglement in this vector is characterized by matrix rank\cite{Schollwoeck2005}. Without treating MPO perturbatively in $\beta$, this entanglement is explicitly determined by the lattice topology and the types of interactions. Nevertheless, the entanglement in MPO is reduced in a simple yet systematic way in this method. It is called the entanglement perturbation theory for a quantum Hamiltonian. The benefit of entanglement reduction in MPO is immediate in that it reduces an eigenvalue problem of rank of $4^N P^2$, which dominates over the GEEs of rank of $2^N P^2$ in terms of computational burden, to that of rank of $\left(3N+1\right)P^2$. It is clear that the bottleneck of simulation of an infinite by $N$ Heisenberg lattice becomes how to efficiently solve GEEs. When $N=14$ and $P=350$ (the parameters causing the most computational burden in the present work), their rank is more than two billion. Yet, solving such a large GEE is not feasible to any existing numerical tool. Thus, we integrate the Jacobi-Davidson method with MPS and MPO to solve the GEEs without explicitly forming them. Details are explained in Sec.\ref{sec:davidson}.

\subsection{\label{subsec:mpsept} Entanglement Entropy and Area Law in Hilbert Space}
It is the entanglement in Hilbert space between an isolated quantum lattice set $\it{I}$ and the surrounding environment $\it{E}$ that is of special interest in the design of a many-body method when using MPS, since the area law\cite{Eisert2010} states that 
\begin{align}
\label{eq:entropy1}
s\left(\it{I}\right)\propto\mid \partial \it{I}\mid
\end{align}
and
\begin{align}
\label{eq:entropy2}
s\left(\it{I}\right)\le 2 log_2\left(P\right)
\end{align}
where $\mid \partial \it{I}\mid$ is the area of boundary $\partial \it{I}$ and $P$ is the MPS rank. The $\it{von}$ $\it{Neumann}$ entanglement entropy $s\left(\it{I}\right)$ is defined as
\begin{align}
\label{eq:entropy}
s\left(\it{I}\right)\equiv -\sum_{i}{\rho_i log_2{\rho_i}}
\end{align}
. Now that the local Hilbert space $H_I\equiv \{\mid \phi_{i}\rangle\}$ is embedded in the complete Hilbert space $H$ to write down a system wave function $\mid \Psi\rangle$, the reduced density matrix is
\begin{align}
\label{eq:densitymatrix0}
\rho_{ij}=Tr_E\langle \phi_{i}\mid \Psi\rangle\langle \Psi\mid \phi_{j}\rangle
\end{align}
. In the case where $\mid \Psi\rangle$ is expressed as a MPS, the reduced density matrix is evaluated as shown in equation \eqref{eq:construction}. See Sec.\ref{sec:reduction} for details. After diagonalizing the reduced density matrix, the entanglement entropy can be readily computed using the diagonal density matrix elements.

\begin{figure}
	\begin{center}
		$\begin{array}{c}
		\includegraphics[width=16.pc]{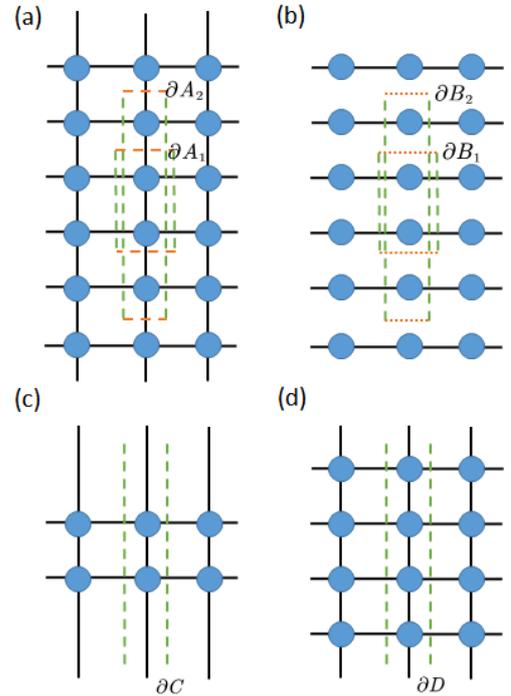}\\
		\end{array}$
		\caption{\label{fig:area_law} Schematic of applicable and inapplicable scenarios of the area law. Applicable: (a) isolations embedded in a large 2D lattice whose sites (circles) interacting in both directions (solid lines). The boundaries in dashed lines are enclosed; (b) the boundary of isolation embedded in a decoupled lattice shows no difference with or without the horizontal dotted lines to enclose them. Inapplicable: from (c) to (d), the width $N$ of the ladder is increasing. They are described by two distinct Hamiltonians. When $N=\infty$ in both (a) and (d), the boundaries are topologically different. The former is single-connected while the latter is disconnected.}\end{center}
\end{figure}

Fig.\ref{fig:area_law}(a) shows the applicable scenario of the area law. An isolated partition increases from $A_1$ to $A_2$, embedded in a given large 2D quantum lattice whose sites (shown as circles) interact in both directions (shown as solid lines). Note that the boundaries $\partial A_1$ and $\partial A_2$ are composed of both vertical and horizontal dashed lines. When the isolation is enlarged, its boundary increases nearly linearly. According to equation \eqref{eq:entropy2}, the MPS rank $P$ which is required to obtain a certain precision for the nearly linearly increasing entanglement entropy should almost exponentially increase. Since the existing many-body methods such as DMRG, the density matrix embedding theory (DMET)\cite{Knizia2012} and the dynamic mean field theory (DMFT)\cite{Metzner1989} all start from or focus on an isolated quantum lattice set surrounded by a large environment as shown in Fig.\ref{fig:area_law}(a), they encounter the same difficulty inherited from the area law of entanglement entropy.

In our method of converting the $N$ lattice sites in the rung into an effective site and then building a MPS on the effective chain, there are two major differences from the scenario shown in Fig.\ref{fig:area_law}(a). First, two lattices of $N_1$ shown in (c) and $N_2$ in (d) of Fig.\ref{fig:area_law} actually have two different Hamiltonians. Second, imagining $N=\infty$ in both (a) and (d) of Fig.\ref{fig:area_law} such that their Hamiltonians are identical, their boundaries $\partial A_{\infty}$ and $\partial D_{\infty}$ are topologically different, because the former is connected and the latter is disconnected. Therefore, the entanglement of isolations $A$ in (a) and $D$ in (d) is different. That is to say, the area law of entanglement entropy is not applicable in our method.

However, Fig.\ref{fig:decoupled} and Fig.\ref{fig:entanglement_decoupled}(c) in Sec.\ref{subsec:decoupled} confirm that the area law coincidently applies when we solve the decoupled spin ladder (Fig.\ref{fig:area_law}(b)) using our method. There are two reasons. First, Hamiltonian $H(n)$ of a decoupled spin ladder of $N=n$ now has the perfect extension property, i.e., $H(n)=nH(1)$ where $H(1)$ is the spin chain Hamiltonian. Second, there is no difference whether or not the boundary is enclosed by the dotted horizontal lines because there is no vertical interaction. 

In fact, the entanglement entropy is crucially controlled by the density matrix of an effective site when treating a quantum ladder by our method. Let's consider the following two ideal cases. 

Case 1. The Hamiltonian has a perfect extensive property, i.e., $H\left(c\right)=cH\left(1\right)$ with $c$ being an integer. Obviously, the decoupled ladder qualifies for this category. In this case, the diagonalized density matrix $\left(\rho\left(c\it{I}\right)\right)$ of isolation $c\it{I}$ is the direct product of $\left(\rho\left(\it{I}\right)\right)=\left(\substack{x,0\\0,y}\right)$ for $c$ times. Then the diagonal element of $\left(\rho\left(c\it{I}\right)\right)$ form a set $\{x^k y^{c-k};k=0,\cdots ,c\}$ with the degeneracy set $\{\left(\substack{c\\k} \right);k=0,\cdots ,c \}$. We have
\begin{align}
\label{eq:entrop3}
s\left(c\it{I}\right)=&-\sum_{k=0}^{c}{\left(\substack{c\\k}\right) x^{c-k}y^k log_2\left(x^{c-k}y^k\right) }\notag\\
=&c\left[\sum_{k=0}^{c-1}{\left(\substack{c-1\\k}\right)x^{c-1-k}y^k}\right] \left(-xlog_2{x}-ylog_2{y}\right)\notag\\
=&c\left(x+y\right)^{c-1}s\left(\it{I}\right)
\end{align}
. Since $x+y=1$,
\begin{align}
\label{eq:entrop4}
s\left(c\it{I}\right)=cs\left(\it{I}\right)
\end{align}
. The area law apparently applies. 

If $x=y$, i.e., the diagonalized density matrix is always equally weighted regardless of the size of $\it{cI}$, there is no dominant element. The MPS rank $P$ that reveals the physical properties such as GS energy to a certain precision should equal to what reveals the entanglement (i.e., the Fock vector configuration in a wave function). Equation \eqref{eq:entropy2} determines that $P$ strictly exponentially increases with $N$ for a demanded energy accuracy. Whereas, the diagonalized density matrix of a decoupled ladder of width $N$ always have dominant elements because $x\ne y$ for a single chain. In turn, our result in Sec.\ref{subsec:decoupled} shows that the increase of $P$ will be slower than an exponential function of $N$ for smaller $N$'s but asymptotically be an exponential function $2^N$ when $N\rightarrow \infty$. 

Case 2. There is a limited number of dominant diagonal elements. We discuss an extreme example:
\begin{align}
	\rho_1=1-2^{-c}\notag\\
	\sum_{i=2}^{\it{R}\left(\it{H}_{c\it{I}}\right)}{\rho_i}=2^{-c}
\end{align}
. Here, $\it{R}\left(\it{H}_{c\it{I}}\right)$ is the Hilbert space rank of $c\it{I}$, and $\rho_{i=2,\cdots,\it{R}\left(\it{H}_{c\it{I}}\right)}$ are equally weighted. In this case, it is easy to show that $s\left(c\it{I}\right)< cs\left(\it{I}\right)$ and that the entanglement entropy saturates when $c$ is large. The area law does not apply. 

In a realistic strongly coupled infinity-by-$N$ spin lattice, our results in Sec.\ref{sec:result} confirm that the area law of entanglement entropy does not apply and the density matrix of an effective site has few, not single, dominant diagonal elements whose number saturates with increasing $N$. 

Letting an effective site have a large local space, our method take the dominant basis vectors into account so as to simulate the physical quantities efficiently with a smaller MPS rank. Packing the entanglement contributed by the dominant basis vectors in a smaller MPS by blocking $N$ quantum lattices in the rung is an implicit entanglement perturbation in the Hilbert space. Meanwhile, space reduction according to the significance of the diagonalized density matrix element is possible as well. We present the details in Sec.\ref{sec:reduction}. For the moment, we discuss in the following three sections other specific properties of the model in this study.   
    
\section{\label{sec:checkerboard} Implementation of Checkerboard Symmetry}
\begin{figure}
	\begin{center}
		$\begin{array}{cc}
		\mbox{(a)}	&\mbox{(b)}   \\
		\includegraphics[width=7.5pc]{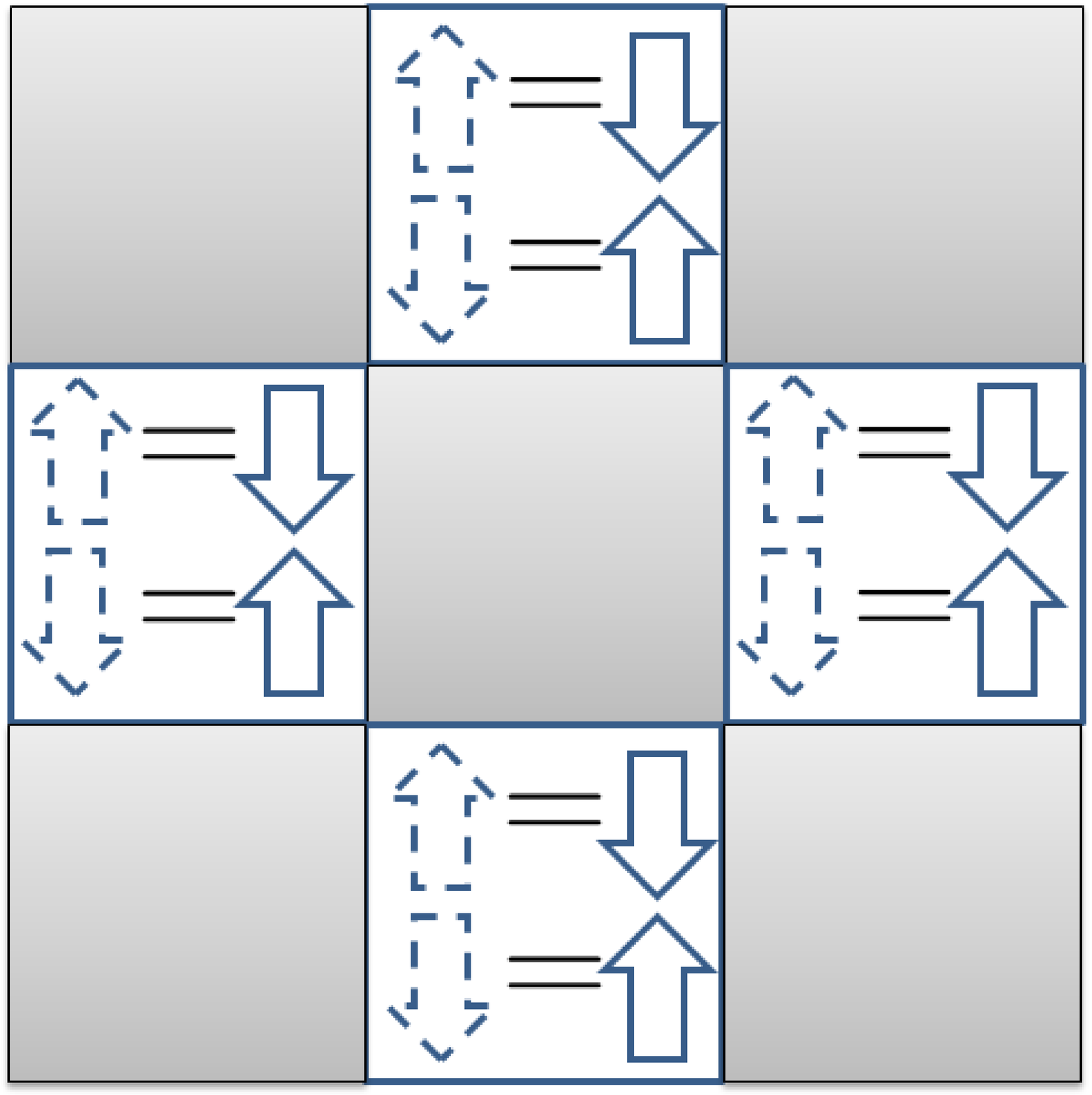} &\includegraphics[width=7.5pc]{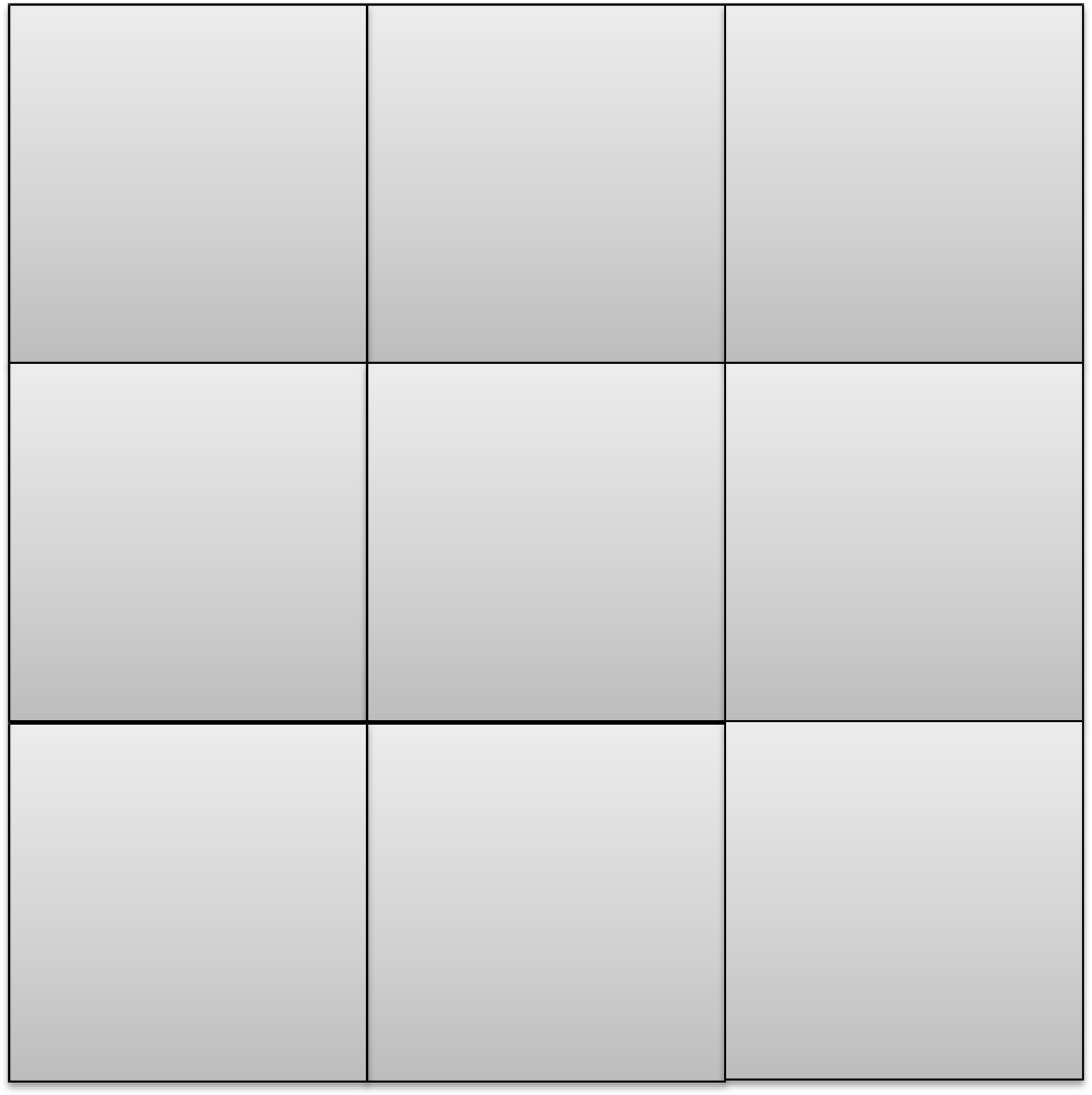}\\
		\end{array}$
		\caption{\label{fig:checkerboard} (a) Owing to the antiferromagnetic nature, the tensors of MPO and MPS have a bipartite checkerboard symmetry in the original spin basis. (b) After applying the checkerboard transformation on any sub-lattice, for instance the blank lattice as shown, those tensors need not be distinguished anymore in the new spin basis.}
	\end{center}
\end{figure} 
So far, we have assumed a bipartite structure for both MPO and MPS according to the antiferromagnetic nature of the studied model. All the formulation can be straightforwardly extended for even more complicated structures. In an opposite limit, we specifically simplify both MPO and MPS to employ a single tensor $\Gamma$ and $\xi$ by a checkerboard transformation applied to a sub-lattice shown in Fig.\ref{fig:checkerboard}(a) as follows, 
\begin{align}
\label{eq:newspin}
\mid \uparrow\prime\rangle\equiv\mid \downarrow\rangle\notag\\
\mid \downarrow\prime\rangle\equiv\mid \uparrow\rangle
\end{align}
. Assuming the second site within a bond is on the transformed sub-lattice, we rewrite $H_{bond}$ as follows, 
\begin{equation}
H_{bond}=\vec{S}_i\cdot\vec{S}_j=S^z_i\cdot S^z_j+\frac{1}{2}\left(S^{+}_i\cdot S^{-}_j+S^{-}_i\cdot S^{+}_j\right)	
\end{equation}
where $S^{+}$ and $S^{-}$ are the spin flip-up and -down operators, respectively. They operate on the transformed site as follows
\begin{align}
\label{eq:checkerboard}
S^z\mid \uparrow\prime\rangle=&-\frac{1}{2}\mid \uparrow\prime\rangle;\hspace{5pc} & S^z\mid \downarrow\prime\rangle=&\frac{1}{2}\mid \downarrow\prime\rangle\notag\\
S^{+}\mid \uparrow\prime\rangle=&\mid \downarrow\prime\rangle; \hspace{5pc} &S^{+}\mid \downarrow\prime\rangle=& 0\notag\\
S^{-}_i\mid \uparrow\prime\rangle=& 0; \hspace{5pc} &S^{-}\mid \downarrow\prime\rangle=& \mid\uparrow\prime\rangle
\end{align}
. Equation (\ref{eq:checkerboard}) is used to rewrite $H_{bond}$ in the checkerboard-transformed basis as
\begin{equation}
\label{eq:checkerboard1}
H_{bond}=-S^z_i\cdot S^z_j+\frac{1}{2}\left(S^{+}_i\cdot S^{\prime +}_j+S^{-}_i\cdot S^{\prime -}_j\right)	
\end{equation}
, where $S^{\prime +}$ ($S^{\prime -}$) has the same matrix representation of $S^{+}$ ($S^{-}$) but flips up (down) the newly defined down (up) spins in equation (\ref{eq:newspin}). After the transformation, we obtain the MPS format in equation (\ref{eq:mps1}) and MPO format in equation (\ref{eq:mpo}). Accordingly, the other related formulation will be simplified and we only present the simplified formulation when they are specifically needed in the remaining formula of this manuscript. Note that a proper sign is needed when a physical quantity is being calculated with the solved wave function in the checkerboard-transformed basis. Also, note that compared with equation \eqref{eq:energy2} before the checkerboard transformation, the GS energy per spin now becomes
\begin{equation}
\label{eq:energy3}
\bar{\epsilon}_0=-\left(N\beta\right)^{-1}ln{\lambda}
\end{equation}
where $\lambda$ is the largest eigenvalue of the single GEE.

\section{\label{sec:davidson} Integration of Jacobi-Davidson eigenvalue solver for GEE with MPO and MPS}
We discuss how to obtain the largest eigenvalue and the corresponding eigenvector by iteratively solving a GEE given by,
\begin{equation}
\label{eq:gee4}
Xx=\lambda Yx
\end{equation}
, where $X$ and $Y$ are $n\times n$ square matrices; $x$ is an eigenvector and $\lambda$ is the corresponding eigenvalue. When only a few eigenvectors are needed (only the largest eigenvalue is needed in this study) and when $n$ is very large, an iterative approach such as the Jacobi-Davidson method\cite{Sleijpen1996} is more desirable than the typical method of factorization. It starts with an initial $n\times m$ matrix $W_0$ whose columns are n-element vectors $w_i,i=1,\cdots,m\ll n$. It is used to transform $X$ and $Y$ to those of rank of $m$ as follows, 
\begin{align}
\label{eq:davidson1}
E_0\equiv W^T_0 X W_0\notag\\
F_0\equiv W_0^T Y W_0
\end{align}
. The following new GEE  
\begin{equation}
\label{eq:davidson2}
E_0y^0=\tau^0F_0y^0
\end{equation}
is much easier to solve for all its eigenvalues $\{\tau^0_i\}$ and the corresponding eigenvectors $\{y^0_i\}$. $\{\tau^0_i\}$ are ordered such that $\tau^0_1>\tau^0_2>\cdots$. Then, a new vector is constructed as follows
\begin{equation}
\label{eq:davidson3}
Q_0=\left(XW_0-\tau^0_1YW_0\right)y^0_1
\end{equation}
. To accelerate the convergence, the vector $Q_0$ is processed to obtain a new vector $w_{m+1}$ as follows,
\begin{equation}
\label{eq:davidson4}
w_{m+1}\left(i\right)=\frac{Q_0\left(i\right)}{\tau^0_1Y\left(i,i\right)-X\left(i,i\right)}; i=1,\cdots,n
\end{equation}
. $w_{m+1}$ is attached to $W_0$ forming an $n\times \left(m+1\right)$ matrix $W_1$. In this procedure, Gram-Schmidt method is used to make $w_{m+1}$ orthonormal to the existing column vectors. On the other hand, the approximated eigenvector corresponding to the largest eigenvalue is 
\begin{equation}
\label{eq:davidson6}
x_0=W_0y_1^0
\end{equation}
. The precision of this approximation is checked by whether $Xx_0$ and $Yx_0$ are parallel. If so, the iteration stops with the solution of the largest eigenvalue $\tau^0_1$ and the corresponding eigenvector $x_0$. If not, we carry on the process with 
\begin{align}
\label{eq:davidson7}
E_1\equiv W_1^T X W_1\notag\\
F_1\equiv W_1^T Y W_1
\end{align}
until $Xx_j$ and $Yx_j$ are parallel at the $j^{th}$ step. Satisfactory convergence is obtained after about 100 steps in this study even for GEE of rank of billions.
 
So far, we introduced the basic steps of Jacobi-Davidson method for GEE. The steps are merely formal because matrices $X$ and $Y$ in fact never appear in the explicit form as shown in equations (\ref{eq:davidson1}), (\ref{eq:davidson3}) and \eqref{eq:davidson7}. They even cannot be generated and stored when their rank is greater than $10^5$, which turns out to be still much behind the requirement to investigate the long-range spin-spin correlation in this study, due to insufficient computational resources. As a workaround, we integrate the aforementioned Jacobi-David method with MPS and MPO, without explicitly forming the left/right matrices of GEE.

After the checkerboard transformation, we only need one MPO tensor $\Gamma$ that is still formulated as in \eqref{eq:omp}. $\Gamma$ is extremely sparse. We only store its nonzero elements in $\{\Omega_j:\mid \Omega_j\mid >0\}$ and the set $\{\left(r_j,s_j,\alpha_j,\gamma_j\right)\}$, each of whose elements is an array of the indices of the $j^{th}$ nonzero element $\Omega_j\equiv\Gamma_{r_js_j,\alpha_j\gamma_j}$. Now, we simplify $X$ and $Y$ in equation \eqref{eq:gee1} as
\begin{align}
\label{eq:newxy}
X_{\lfloor r,\alpha_1,\gamma_1\rfloor ,\lfloor s,\alpha_3,\gamma_3\rfloor}&=v_{\alpha_1\alpha_2\alpha_3}\Gamma_{rs,\alpha_2\gamma_2} u_{\gamma_1\gamma_2\gamma_3}\notag\\
Y_{\lfloor r,\alpha_1,\gamma_1\rfloor ,\lfloor s,\alpha_3,\gamma_3\rfloor}&={v'}_{\alpha_1\alpha_3}\delta_{r,s} {u'}_{\gamma_1\gamma_3}
\end{align}
. These are dense matrices. The notion of effective site's label is no longer needed since the lattice is translationally symmetric after the checkerboard transformation. 

Except explicitly evaluating equation \eqref{eq:newxy} for the diagonal element of $X$ and $Y$ that are needed in equation (\ref{eq:davidson4}), the matrix-vector multiplication only in which $X$ and $Y$ explicitly participate an operation (equations (\ref{eq:davidson1}), (\ref{eq:davidson3}) and \eqref{eq:davidson7}) can be evaluated as follows,
\begin{equation}
\label{eq:davidson8}
z_{\lfloor r,\alpha_1,\gamma_1\rfloor}=\sum_{j}{v_{\alpha_1\alpha_j\alpha_3}\Gamma_{r_js_j,\alpha_j\gamma_j} u_{\gamma_1\gamma_j\gamma_3} w_{\lfloor s_j,\alpha_3\gamma_3\rfloor}}
\end{equation}
, where $r$ is dynamically updated by $r_{j}$ during the summation over $j$. We further rewrite equation (\ref{eq:davidson8}) as follows     
\begin{equation}
\label{eq:davidson9}
z_{\lfloor r,\alpha_1,\gamma_1\rfloor}=\sum_{j}{\Gamma_{r_js_j,\alpha_j\gamma_j}\left(\pi^{\alpha_j}\cdot\varpi^{s_j}\cdot\rho^{\gamma_j}\right)_{\alpha_1 \gamma_1}}
\end{equation}
. In the above, three newly defined matrices participate in a chain of product to yield a resultant matrix in the parentheses. Equation \eqref{eq:davidson9} is equivalent to update $z$ with the resultant matrix for $L$ times, where $L$ is the number of nonzero elements of $\Gamma$ and these elements are the updating coefficients. Those three matrices in parentheses are defined using the tensors appearing in \eqref{eq:davidson8} as
\begin{align}
\label{eq:davidson10}
\left(\pi^k\right)_{ij}\equiv v_{ikj}; \hspace{1pc}\left(\varpi^k\right)_{ij}\equiv w_{\lfloor kij\rfloor}; \hspace{1pc}\left(\rho^k\right)_{ij}\equiv u_{jki}
\end{align}
. Note the difference of in the right hand side of the first and third equations in (\ref{eq:davidson10}), which reflects the sandwich structure constructed by the MPS wave function, its conjugate and the MPO. The transformation from expression of matrix-vector multiplication in equation (\ref{eq:davidson8}) to that in (\ref{eq:davidson9}) is crucial in that the summations are broken into successive matrix product of matrices, which reduces the computational cost by a few orders of matrix rank $P$. Again, the method that associates many loops of index-summation by matrix product, helps reduce the computational burden when solving GEE in the presence of MPS and MPO. 

Meanwhile, the SVDs in equation \eqref{eq:ABCD} are simplified after the checkerboard transformation as  
\begin{align}
\label{eq:ABCD1}
A=& u\Lambda v\notag\\
C=& u'\Delta v'
\end{align}
where
\begin{align}
\label{eq:relabel2}
A_{\alpha\equiv\lfloor\alpha_1\alpha_2\alpha_3\rfloor ,\gamma\equiv\lfloor\gamma_1\gamma_2\gamma_3\rfloor}\equiv &\xi_{r,\alpha_1\gamma_1}\Gamma_{rs,\alpha_2\gamma_2}\xi_{s,\alpha_3\gamma_3}\notag\\
C_{\alpha\equiv\lfloor\alpha_1\alpha_3\rfloor ,\gamma\equiv\lfloor\gamma_1\gamma_3\rfloor}\equiv &\xi_{r,\alpha_1\gamma_1}\xi_{r,\alpha_3\gamma_3}
\end{align}
which have the rank of $T=\left(3N+1\right)P^2$ and $O=P^2$, respectively. Both of them will be very large in some cases. For instance, the largest $T$ value encountered in this study is $1.0\times 10^8$ when $P=2000$ for $N=8$. It is not accessible by ordinary SVD solvers. In this study, SVDs are solved by the power method without forming $A$ and $C$. The matrix-vector multiplication is transformed into a successive matrix product similar to equation \eqref{eq:davidson9}. 

The composite matrices $A$, $B$ defined in equation \eqref{eq:relabel} have rank $\left(3N+1\right)P^2$; $C$ and $D$ defined in equation \eqref{eq:relabel1} have rank $P^2$. The composite-matrix-vector multiplication is the time-controlling factor in the Davidson method. It is decomposed to a chain of lower-ranked matrix-vector products after decomposing the composite matrix as in equation \eqref{eq:davidson9}. The lower rank is $P$. Thus, the time consumption in iqEPT is bilinear with both $P^{2.5}$ (if using a Lapack routine) and $2^N$ (which determines the number of nonzero elements in $\Gamma$). In the largest simulation scale of this work, it took $2\times 10^4$ seconds to solve the GEE in a series of iterations for $N=10$ and $P=1400$, using $6$ Dual-Intel-Xeon nodes each of which has $20$ cores and $256$ GB memory installed. It took $50$ iterations to obtain the converged data for that set of $N$ and $P$.      
 
\section{\label{sec:correlation} spin-spin correlation and local magnetization}
The spin-spin correlation is defined as $C_r\equiv\langle S^z_{\left(i,j\right)} S^z_{\left(i+r,j\right)}\rangle$ where the operators are separated by $r$ spins in LD of an infinity-by-$N$ lattice. Without loss of generality we set $j=1$. After converting the lattice into a chain of effective lattice sites, the operators are redefined as
\begin{align}
\label{eq:operator}
S^z_{i,\text{eff}}\equiv S^z_{\left(i,1\right)}\otimes I_{\left(i,2\right)}\otimes\cdots\otimes I_{\left(i,N\right)}
\end{align}
, where $I_{\left(i,m\right)}$ is an identity operator operating on $\left(i,m\right)^{th}$ physical lattice site.

It is straightforward to construct a tensor $\Theta$ for $S^z_{i,\text{eff}}$. After implementing the checkerboard transformation, the same tensor also applies to $S^z_{i+r,\text{eff}}$. Therefore,
\begin{widetext}
\begin{align}
\label{eq:correlation}
C_r=\frac{tr\left[\cdots\left(\xi_{s^{i},\alpha_i\alpha_{i+1}}\Theta_{s^{i}r^{i}}\xi_{r^{i},\gamma_i\gamma_{i+1}}\right)\cdot\left(\xi_{s^{i+1},\alpha_{i+1}\alpha_{i+2}}\xi_{s^{i+1},\gamma_{i+1}\gamma_{i+2}}\right)\cdots
\left(\xi_{s^{i+r},\alpha_{i+r}\alpha_{i+r+1}}\Theta_{s^{i+r}r^{i+r}}\xi_{r^{i+r},\gamma_{i+r}\gamma_{i+r+1}}\right)\cdots\right]}{tr\left[\cdots \left(\xi_{s^{i},\alpha_{i}\alpha_{i+1}}\xi_{s^{i},\gamma_{i}\gamma_{i+1}}\right)\cdot \left(\xi_{s^{i+1},\alpha_{i+1}\alpha_{i+2}}\xi_{s^{i+1},\gamma_{i+1}\gamma_{i+2}}\right)  \cdots\right]}
\end{align}
\end{widetext}
We convert the quantities in the first and second parentheses in the numerator of equation (\ref{eq:correlation}) to matrices $G$ and $B$ as we did in equation \eqref{eq:relabel} in Sec.\ref{sec:mps}. Equation (\ref{eq:correlation}) becomes
\begin{align}
\label{eq:correlation1}
C_r=\lim_{M\rightarrow \infty}{\frac{tr\left(G\cdot B^r\cdot G\cdot B^{M-r-2}\right)}{tr\left(B^M\right)}}
\end{align}
. Since equation (\ref{eq:correlation1}) only yields positive values, a proper sign should be given to $C(r)$ according to even or odd $r$, to reflect the checkerboard transformation. We singular-value-decompose $B$ as $B=\mu \varrho \nu$. $\{\varrho_i\}$ is sorted in descending absolute magnitude. Equation \eqref{eq:correlation1} leads to 
\begin{align}
\label{eq:correlation2}
C_r=\frac{tr\left({\nu}^T_1\cdot G\cdot B^r\cdot G\cdot{\mu}_1\right)}{\varrho_1^{r+2}}
\end{align}
. When $r\rightarrow \infty$, we have
\begin{align}
\label{eq:correlation3}
C_{\infty}=\frac{F_1^2}{\varrho_1^2}
\end{align}
where
\begin{align}
\label{eq:F}
F_1\equiv {\nu}^T_1\cdot G\cdot{\mu}_1
\end{align}
. The numerator may be zero or nonzero, giving disorder including quasi-long-range-order (QLRO) or order of the lattice. Moreover, it is straightforward that the local magnetization is
\begin{align}
 \label{eq:correlation4}
 \bar{M}\equiv\langle S^z\rangle=\lim_{M\rightarrow \infty}{\frac{tr\left(G\cdot B^{M-1}\right)}{tr\left(B^M\right)}}=\frac{F_1}{\varrho_1}
\end{align}
. Thus, we arrive at a theorem that the spin-spin correlations after infinite spin-spin separation is square of the local magnetization, 
\begin{align}
 \label{eq:correlation5}
 C_{\infty}=\bar{M}^2
\end{align}
. This theorem is a useful supplement to the commonly used definitions relating spin-spin correlations to staggered magnetization\cite{Manousakis1991} though they are not universally agreed upon\cite{Kaplan1989}.  

A new quantity
\begin{align}
\label{eq:tau}
\tau_r\equiv Ln\left(LnC_r-LnC_{r+1}\right)
\end{align}
is proven in Sec.\ref{subsec:coupled} to be useful. Although only the largest eigenvalue of $B$ determines the asymptotic $C_\infty$, the first eigenvalue $\varrho_{\bar{k}}$ that has significant non-vanishing
 \begin{align}
\label{eq:Fk}
F_{\bar{k}}\equiv {\nu}^T_1\cdot G\cdot{\mu}_{\bar{k}}={\nu}^T_{\bar{k}}\cdot G\cdot{\mu}_1
\end{align}
is also important to determine how $C_r$ approaches $C_\infty$ asymptotically. Since
 \begin{align}
\label{eq:correlation6}
C_r\approx \frac{F_1^2}{\varrho_1^2}+\frac{F_{\bar{k}}^2}{\varrho_1^2}\left(\frac{\varrho_{\bar{k}}}{\varrho_1}\right)^{r-1}
\end{align}
, we have 
 \begin{align}
\label{eq:correlation7}
\tau_r \approx \left(r-1\right)\left(Ln\varrho_{\bar{k}}-Ln\varrho_1\right)+\it{f}
\end{align}
where
 \begin{align}
\label{eq:correlation8}
\it{f}\equiv 2\left(LnF_{\bar{k}}-LnF_1\right)+Ln\left(1-\varrho_{\bar{k}}/\varrho_1\right)
\end{align}
. Equation \eqref{eq:correlation4} tells that $F_1$ is not vanishing for an ordered lattice. However, $F_1$ neither vanishes when the simulation of staggered magnetization $\bar{M}$ does not reach the convergence of zero yet for a disordered lattice. In both cases, equation \eqref{eq:correlation7} reads that $\tau_r$ is linear with $r$. 

There are three different scenarios of $\tau_r$ versus $r$ after $F_1$ is converged with $P$. The first two scenarios are,

Case 1. $F_1\ne 0$. The slope of $\tau_r$ versus $r$ is a nonzero constant $p_0 \equiv Ln\varrho_2-Ln\varrho_1$. Then, $LnC_r$ hence $C_r$ becomes constant for large $r$. The lattice is ordered.

Case 2. $F_1=0$. Equation \eqref{eq:correlation7} loses definition. IF $\varrho_{\bar{k}}$ is significantly smaller than $\varrho_1$, Equation \eqref{eq:correlation6} reads as
\begin{align}
\label{eq:correlation9}
C_r\approx \frac{F_{\bar{k}}^2}{\varrho_1^2}\left(\frac{\varrho_{\bar{k}}}{\varrho_1}\right)^{r-1}
\end{align}
. The slope of $\tau_r$ versus $r$ is zero. $C_r$ exponentially decays with $r$. The lattice is disordered and gapped.

In the following third scenario, $\tau_r$ behaves uniquely.  

Case 3. The first few largest eigenvalues are almost degenerate with $\varrho_1$. But, they have less significant $F_j$. They give very small correlation-contribution that slowly decays. The eigenvalues which have significant $F_j$'s are however definitely smaller than $\varrho_1$. They give correlation-contribution that is large for small $r$ but exponentially decays. All summed up, the resultant correlation function shows a power-law decay in a large range of $r$. $\tau_r$ is linear with $Ln\left(r\right)$ instead of $r$. The lattice is QLRO. See Appendix.\ref{sec:appendix} for an example of spin chain.

\section{\label{sec:reduction} space reduction in matrix product state}
  
The use of density matrix is crucial in simulating quantum lattice model in that it guides reduction of Hilbert space/subspace which exponentially increases with respect to the system/subsystem size. Given a bipartite structure $A$ and $E$ of a system, the system wave function $\mid \Psi\rangle$ is an entangled quantity composed of basis vectors from subspaces $\{\mid \phi_i\rangle\}$ for $A$ and $\{\mid \psi_i\rangle\}$ for $E$, 
\begin{equation}
\label{eq:systemwavefunction}
\mid \Psi\rangle=\sum_{i,j}{X_{ij}\mid \phi_{i}\rangle\mid \psi_{j}\rangle}
\end{equation}
where $X$ is a tensor entangling $A$ and $E$. The density operator for a subspace, say $\{\mid \phi_i\rangle\}$, is defined as $\hat{\rho}\equiv Tr_E\mid \Psi\rangle\langle \Psi\mid$, where $Tr_E$ means that the degree of freedom in subsystem $E$ is traced out. Its matrix representation is 
\begin{align}
\label{eq:densitymatrix}
\left(\rho_{ij}\right)\equiv \langle \phi_i\mid \hat{\rho}\mid \phi_j\rangle=X_{ik}X^*_{jk}
\end{align}
. Note that normalization of $\mid \Psi\rangle$ implies $tr\rho=1$. 

The system wave function can be reconstructed as $\mid \Psi^{\prime}\rangle$ using a reduced space $\{\mid\theta_i\rangle\}$ consisted of $M$ basis vectors for $A$ along with the unaltered space $\{\mid \psi_i \rangle\}$ for $E$. The density matrix built in $\{\mid \phi_i\rangle\}$ is used to make the residual vector $\mid R\rangle\equiv \mid \Psi\rangle-\mid \Psi^{\prime}\rangle$ have a minimum norm\cite{Schollwoeck2005,White1992,White1993}. Explicitly, the density matrix is diagonalized and only $M$ eigenvectors $\{v_i;i=1,\cdots ,M\}$ need to be retained. They correspond to the most significant $M$ diagonal elements $\{\eta_i\}$. New basis vectors are constructed as
\begin{align}
\label{eq:basis_transformation}
\mid \theta_i\rangle=v_i\left(k\right)\mid \phi_k\rangle
\end{align}
where $v_i\left(k\right)$ is the $k^{th}$ element of the eigenvector $v_i$. And, $\mid \Psi^{\prime}\rangle$ is constructed as
\begin{align}
\label{eq:reduced_function}
\mid \Psi^{\prime}\rangle=Y_{ij}\mid \theta_{i}\rangle\mid \psi_{j}\rangle
\end{align}
where
\begin{align}
\label{eq:function_transformation}
Y_{ij}=v_i\left(k\right)X_{kj}
\end{align}
. If the formal reduction is unitary (zero reduction), substituting equations \eqref{eq:basis_transformation} and \eqref{eq:function_transformation} into equation \eqref{eq:reduced_function} restores the wave function in equation \eqref{eq:systemwavefunction}. When the truncation of space of $A$ takes place, i.e, when the eigenvector matrix kept is rectangular, hence no longer unitary, we have
\begin{equation}
\label{eq:restore_function}
\mid \Psi^{\prime}\rangle = X_{kj}v_i(k)v_i(l)\mid \phi_{l}\rangle\mid \psi_{j}\rangle = \mid \Psi\rangle-\mid R\rangle
\end{equation} 
where
\begin{equation}
\label{eq:residual}
\mid R\rangle=X_{kj}\Delta_{kl}\mid \phi_{l}\rangle\mid \psi_{j}\rangle
\end{equation}
. Here, $\Delta_{ij}=\sum_{k=M+1}^{n}v_i\left(k\right)v_j\left(k\right)$. It is straightforward to show $||\mid R\rangle||^2=\sum_{i=M+1}^{n}{\eta_i}$.

Following the line of local space reduction, DMRG uses the density matrix to keep a fixed amount of transformed basis vectors for an enlarged part of system. Meanwhile, DMET provides an alternative to DMFT, using the density matrix to improve the impurity state of a fragment embedded in background.   

We implement the density matrix in a different way where it is used to reduce spaces in MPS. Dividing a quantum lattice into $L$ blocks each of which contains $N$ physical sites, a MPS is built as follows
\begin{align}
\label{eq:mps2}
\mid \Psi\rangle=\sum_{\cdots r^i\cdots r^L}{tr\left(\cdots \xi_{r^{i-1}}^{i-1}\cdot \xi_{r^i}^i\cdots\right)\cdots\mid \phi_{r^{i-1}}^{i-1}\rangle\mid \phi_{r^{i}}^{i}\rangle\cdots} 
\end{align} 
where each MPS tensor $\xi^i$ is associated with a block, say, the $i^{th}$ block. Different from equation \eqref{eq:mps1}, the MPS in equation \eqref{eq:mps2} has a more general form. Its tensor does not necessarily have a unit cell structure and the space index $r^i$ runs from $1$ to $Q\equiv R^N$ meaning that each of the $N$ physical sites in a block has a general space rank $R$.

The computational burden of variational optimization of each MPS tensor is determined by both bonding rank $P$ and space rank $Q$. One needs a tractable strategy to balance between choices of $P$ and $Q$. Choosing blocks that contain more physical sites has the following benefits. First, there are fewer tensors to solve. Second, it uses smaller $P$ to achieve the same precision. In the extreme case when a block contains the whole system, one just needs a MPS tensor of rank $1$ to precisely represent the system wave function. However, as $Q$ increases exponentially with $N$ to exclude possibility of building MPS on a block containing the whole system, one still needs to solve multiple MPS tensors, while the same computational resource only allows smaller $P$ when $N$ is larger. 

We propose a scheme to overcome this difficulty when building MPS on a blocked quantum lattice. A MPS ranked $P_1$ in the original spaces, $\mid \Psi\rangle_{\perp P= P_1}$, is used to construct density matrix for each block, say the $i^{th}$ block, as 
\begin{widetext}
	\begin{align}
	\label{eq:construction}
	\rho_{ab}^{i}=tr\left[\cdots\left(\xi^{i-1}_{r^{i-1},\alpha_1\gamma_1}\xi^{i-1}_{r^{i-1},\alpha_2\gamma_2}\right)\left(\xi^i_{a,\gamma_1\eta_1}\xi^i_{b,\gamma_2\eta_2}\right)\left(\xi^{i+1}_{r^{i+1},\eta_1\theta_1}\xi^{i+1}_{r^{i+1},\eta_2\theta_2}\right)\cdots\right]
	\end{align}
\end{widetext}
. In $\xi^i_{a,\gamma_1\eta_1}$, $a$ denotes the space index and the bond indices $\gamma_1$ and $\eta_1$ are explicitly shown here. Note that the density matrix element in equation \eqref{eq:construction} should be adjusted according to the MPS normalization. Density matrices will be constructed for all $L$ tensors (blocks) and are diagonalized simultaneously. For each block, only eigenvectors corresponding to the most significant $M$ diagonal elements are used to transform the space $\{\mid \phi^i_j\rangle\}$ to the smaller one $\{\mid \theta^i_j \rangle\}$ according to equation \eqref{eq:basis_transformation}, where the superscript refers to the block's label.

Note that MPS in the reduced space set $\{H^{i,\prime}\equiv\{\theta^i_j\}\}$ for $P\le P_1$, $\mid \Psi^{\prime}\rangle_{\perp P\le P_1}$, can be reconstructed from the existing MPS $\mid \Psi\rangle_{\perp P\le P_1}$ as 
\begin{align}
\label{eq:newmps}
\mid \Psi^{\prime}\rangle=\sum_{\cdots s^i\cdots s^L}{tr\left(\cdots \kappa_{s^{i-1}}^{i-1}\cdot \kappa_{s^i}^i\cdots\right)\cdots\mid \theta_{s^{i-1}}^{i-1}\rangle\mid \theta_{s^{i}}^{i}\rangle\cdots} 
\end{align}
where 
\begin{align}
\label{eq:newmpstensor}
\kappa^i_{a}=v_a\left(b\right)\xi^i_{b}
\end{align}
. $\mid \Psi^{\prime}\rangle_{\perp P> P_1}$ is then variationally determined. Or, the MPS in reduced spaces can be variationally determined for all the range of $P$. The variational solution is described in Sec.\ref{sec:ept}. In both ways, since $\{H^{i,\prime}\}$ has a smaller rank $Q^{\prime}<Q$, the same computational resource now allows larger $P$, yielding better accuracy in turn.

For the spin ladder studied in this work, we showed that the linking complexity between the building units of MPO is reduced to a linear dependence on $N$ and the large-size GEE is efficiently handled by integrating the Jacobi-Davidson method with MPS and MPO. The only remaining factor that gains exponential complexity with increasing lattice width is the space rank $2^N$ of an effective site. This rank has been crucial in handling even larger lattices or different complex scenarios by iqEPT. This exponential complexity is now systematically overcome by the space reduction in MPS.

Meanwhile, the density matrix can be used to reveal the system properties since it determines which kind of basis vector contributes the most in the GS wave function. To this end, we define the following quantity to reveal the spontaneous spin rotational symmetry breaking, 
\begin{align}
\label{eq:neworder}
\bar{S^z}_j  \equiv \langle \theta_j^i \mid \sum_{k=1}^{N}{\left(-1\right)^k S^z_{\left(i,k\right)}} \mid \theta_j^i\rangle
\end{align} 
where $\mid \theta^i_j\rangle$ is the $j^{th}$ new basis vector in the space (only unitarily transformed, not necessarily reduced) of the $i^{th}$ effective site. $\left(i,k\right)$ is the 2D coordinates of a physical site. For a wave function having the broken symmetry, one spin configuration is no longer equivalent to its up-side-down counterpart. One is expected to have a larger amplitude than the other in GS. It leads to nonzero values of $\bar{S^z}_j$ with the same sign for those most significant $j's$. By contrast, this quantity is zero when there is no spin rotational symmetry breaking. 

\section{\label{sec:result}results}
\subsection{\label{subsec:decoupled} Benchmark and comparison}
\begin{table*}[ht]
	\centering	
	\begin{tabular}{c|cccc|cc}
		\hline
		& \multicolumn{4}{c|}{iqEPT} & DMRG&\\
		\cline{2-5} 
		N & $P^{\prime}$ & $\epsilon_{P^{\prime}}$ & extrapolation & $\Delta$ & extrapolation & MC Loop\\
		\hline 
		$2$ & $40$   & $-0.578043$ & $-0.578043$ &  					&  $-0.578043$	& $-0.57802$\\
		$3$ & $500$  & $-0.600538$ & $-0.600538$ &  					& $-0.600537$ 	& $-0.60063$\\
		$4$ & $640$  & $-0.618567$ & $-0.618567$ & 						& $-0.618566$ 	& $-0.61873$\\
		$5$ & $1200$ & $-0.627781$ & $-0.627787$ & $9.6\times 10^{-6}$	& $-0.62776$ 	& $-0.62784$\\
		$6$ & $1600$ & $-0.634681$ & $-0.634690$ & $1.4\times 10^{-5}$	& $-0.6346$ 	& $-0.635(1)$\\						
		\hline
	\end{tabular}
	\caption{\label{table:compare}Comparison of GS energy per site among iqEPT, DMRG and MC loop algorithm for spin ladders with open boundary condition imposed in the rung.}
\end{table*}
\begin{figure}
	\begin{center}
		$\begin{array}{c}
		\mbox{(a)}\\		
		\includegraphics[width=18pc]{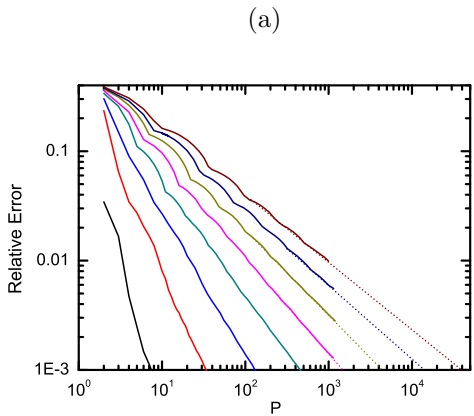} \\
		\mbox{(b)}\\		
		\includegraphics[width=18pc]{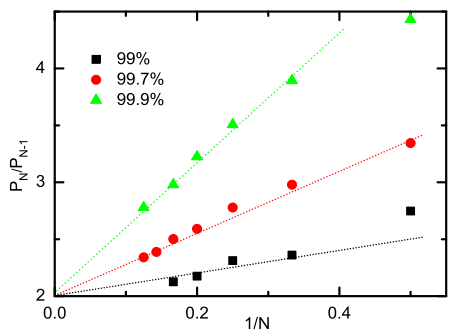}\\  		  
		\end{array}$
		\caption{\label{fig:decoupled} (a) Relative error versus MPS rank $P$ for decoupled ladders of $N=1, 2, 3, 4, 5, 6, 7$ and $8$, from left hand side to right hand side. (b) Ratio of $P_N$ (for $N$) to $P_{N-1}$ (for $N-1$), with respect to $\frac{1}{N}$. At both $P_N$ and $P_{N-1}$ the same accuracy is obtained. The comparison is made for accuracies $99.9\%$, $99.7\%$, and $99\%$ from top to bottom. The tendency of $\frac{P_N}{P_{N-1}}\rightarrow 2$ for $\frac{1}{N}\rightarrow 0$ implies that the MPS rank $P$ will asymptotically increase as an exponential function $2^N$.}  
	\end{center}
\end{figure}
\begin{figure}
	\begin{center}
		$\begin{array}{c}
		\mbox{(a)}\\
		\includegraphics[width=18.pc]{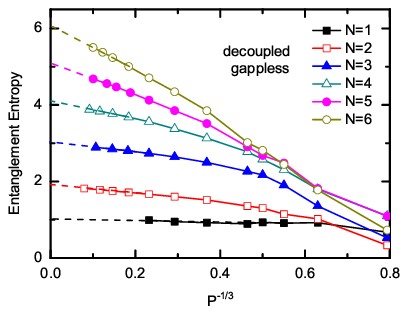}\\	
		\mbox{(b)}\\
		\includegraphics[width=18.pc]{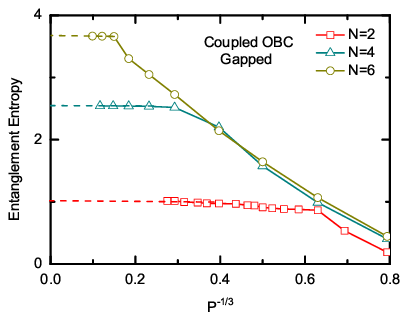}\\			
		\mbox{(c)}\\
		\includegraphics[width=18.pc]{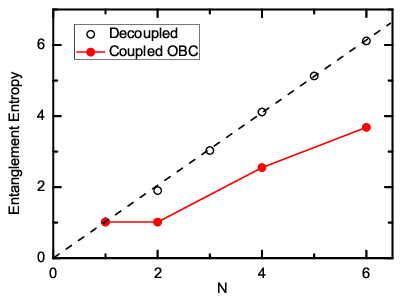}\\
		\end{array}$
		\caption{\label{fig:entanglement_decoupled} Entanglement entropy of a rung versus $P^{-1/3}$ for (a) decoupled ladders and (b) coupled ladders with OBC in the rung. (c) Entanglement entropy versus $N$. Open and solid circles represent coupled and decoupled ladders, respectively.}
	\end{center}
\end{figure} 
\begin{figure}
	\begin{center}
		$\begin{array}{c}
		\mbox{(a)}\\				
		\includegraphics[width=18pc]{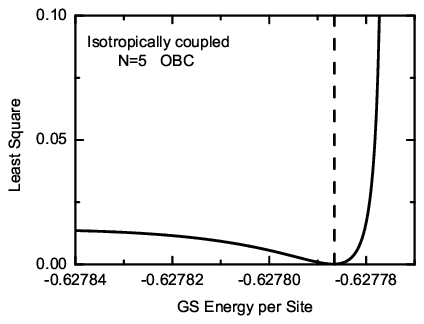}\\  
		\mbox{(b)}\\		
		\includegraphics[width=18pc]{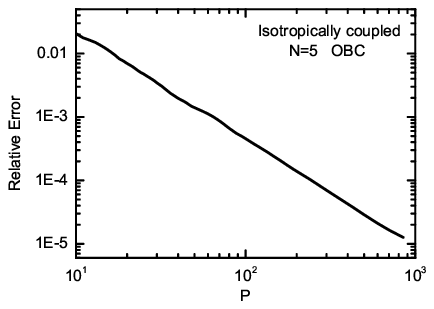}			 
		\end{array}$
		\caption{\label{fig:loglogfit} (a) Least square versus GS energy per site and (b) log-log view of relative error versus MPS rank $P$ for an isotropically coupled ladder of $N=5$ with OBC in the rung. The minimum of curve in (a) gives the extrapolation at $P\rightarrow\infty$ for GS Energy per site needed to calculate the relative error in (b).}
	\end{center}
\end{figure} 
\begin{figure}
	\begin{center}
		$\begin{array}{c}
		\mbox{(a)}\\
		\includegraphics[width=18pc]{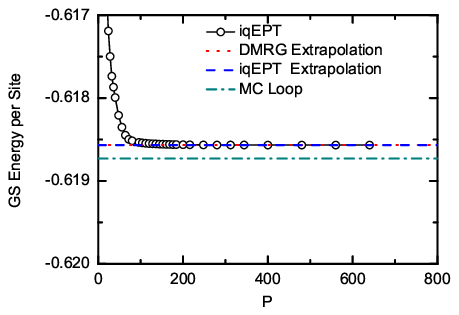}\\
		\mbox{(b)}\\
		\includegraphics[width=18pc]{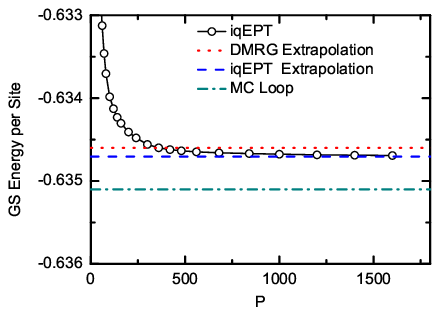}\\   
		\end{array}$
		\caption{\label{fig:compare} Comparison of GS energy per site for spin ladders of $N$ (a) $=4$ and (b) $=6$ both in the energy scale of $0.003$, by iqEPT, DMRG and MC loop algorithm. The discrepancy between the short dotted DMRG extrapolation and the converged iqEPT result, that overlaps with its dashed extrapolation, increases by two orders in magnitude when $N$ increases from $4$ to $6$. Meanwhile, MC loop algorithm result is not reliable in the energy scale shown.}
	\end{center}
\end{figure} 
In order to check the correctness of our algorithm, we benchmark the method on decoupled spin-$\frac{1}{2}$ ladders for various width $N$'s. Regardless of $N$, the ground state energy per site $\epsilon_0$ should be equal to the exact value $-0.443147$ by Bethe ansatz\cite{Bethe1931} of a single spin chain. Previously, the results for $N=1$ along a similar line of reasoning were reported\cite{Wang2012} to agree with the exact results. There was no need to utilize the entanglement reduction in MPO described in Sec.\ref{sec:mpoept} and didn't apply the algorithm extensions presented in this work. For $N=2$, our extended algorithm reproduced the exact energy at $P=2000$, proving its correctness.  

Fig.\ref{fig:decoupled}(a) shows a linear log-log relationship between the error of iqEPT data, relative to the exact value, and the MPS rank $P$. Fig.\ref{fig:decoupled}(b) further shows that the ratio of $P_N$ (for $N$) to $P_{N-1}$ (for $N-1$) approaches $2$ when $1/N\rightarrow 0$. This indicates that $P$ will asymptotically increase as an exponential function $2^N$ when $N\rightarrow \infty$. This increase is very rapid. For example, the MPS rank needed to obtain an energy accuracy of $99.99\%$ is about $6\times 10^5$ for $N=6$. It is clear that treating a decoupled spin ladder by iqEPT is inefficient.

We compute the entanglement entropy according to equation \eqref{eq:entropy} after diagonalizing the density matrix of a rung obtained in equation \eqref{eq:construction}. Fig.\ref{fig:entanglement_decoupled}(a) shows that it has a linear dependence on $P^{-1/3}$. It can be used to make reliable extrapolations used in (c). The linear dependence of the entanglement entropy versus $N$, shown as open circles, confirms our prediction made in Sec.\ref{subsec:mpsept} that the area law of entanglement entropy coincidently applies to decoupled ladders in our method. Note that, the convergence of entropy is continuous in (a) for the gapless decoupled ladders. In what follows, however, the scenario changes drastically for an isotropically coupled spin ladder with either PBC or open boundary condition (OBC) imposed in the rung. First is the ladder with OBC in the rung for $N$ up to $6$, to directly compare with the existing methods in the literature. 

The solid circles in Fig.\ref{fig:entanglement_decoupled}(c) show that the area law does not apply to the coupled ladder (see Sec.\ref{subsec:mpsept} for explanation). And, the sudden convergence of entanglement entropy in (b) shows that the coupled ladder with OBC in the rung is gapped\cite{Eisert2010} for $N=2, 4$ and $6$. However, the gap does not exponentially decay with increasing $N$ because the entanglement entropy will otherwise be linear\cite{Irani2010,Gottesman2010,Eisert2010} with $N$. This observation does not yet violate NLSM's prediction on the exponential decay of the gap because this prediction only applies to the coupled ladder with PBC in the rung. To this end, the entanglement entropy in the target model, i.e., the coupled ladder with PBC, is computed. Fig.\ref{fig:entanglement_coupled}(c) partially confirms the prediction for $N=2, 4$ and $6$, as the entropy segment within this interval of $N$ is indeed linear. Nevertheless, the saturation starting from $N=8$ suggests that NLSM's prediction does not apply to larger $N$'s. See Sec.\ref{subsec:coupled} for details. For the moment, we continue to use the model with OBC to compare with the existing methods.  

Since there is no exact result to calculate the relative error directly, we extrapolate the asymptotic energy $\bar{\epsilon}\equiv \epsilon_{P\rightarrow\infty}$ to obtain the relative error. Assuming a power relationship between the relative error and $P$ in iqEPT for isotropically coupled ladders, the parameterized least square is defined as follows
\begin{align}
	\label{eq:fitting}
	\it{l}\left(\bar{\epsilon}\right)\equiv \sum_{i}\left(b_1 logP_i+b_2-log\frac{\bar{\epsilon}-\epsilon_{P_i}}{\bar{\epsilon}}\right)^2\notag\\
\end{align}
where
\begin{align}
	\label{eq:fitting1}
b_1=&\frac{m\sum_i{logP_i log\frac{\bar{\epsilon}-\epsilon_{P_i}}{\bar{\epsilon}}}-\sum_i{logP_i}\sum_i{log\frac{\bar{\epsilon}-\epsilon_{P_i}}{\bar{\epsilon}}}  }{m\sum_i{\left(logP_i\right)^2}-\left(\sum_i{logP_i}\right)^2}\notag\\
b_2=&\frac{1}{m}\left(\sum_i{log\frac{\bar{\epsilon}-\epsilon_{P_i}}{\bar{\epsilon}}}-b_1\sum_i{logP_i}\right)
\end{align}
. Minimizing equation \eqref{eq:fitting} gives the optimal extrapolation $\bar{\epsilon}$. Fig.\ref{fig:loglogfit}(a) shows an example of such extrapolation for $N=5$. Fig.\ref{fig:loglogfit}(b) confirms the assumed power relationship, making the extrapolation self-consistent.

Extrapolations are made for all ladders in comparison. Fig.\ref{fig:compare} shows that the GS energy for $N=4$ and $6$ have converged in the scale shown. Table.\ref{table:compare} compares results by iqEPT, DMRG\cite{Ramos2014} and MC loop algorithm\cite{Frischmuth1996}. Data given by DMRG in \cite{Ramos2014} is not the actually obtained data but its extrapolation after two loops of scaling which vary both the finite ladder length and the number of kept diagonal elements in density matrix. Both Fig.\ref{fig:compare} and Table.\ref{table:compare} show that the discrepancy between iqEPT results (including extrapolation) and DMRG extrapolation rapidly increases when $N$ increases from $4$ to $6$. Explicitly, it is $1.62\times 10^{-6}$ ($1.62\times 10^{-6}$) for $N=4$ and $1.28 \times 10^{-4}$ ($1.42 \times 10^{-4}$) for $N=6$, increasing by two orders in magnitude. For larger $N$, the discrepancy is expected to be progressively larger. As the next subsection will show, for the ladder of interest which has PBC imposed in both directions, the relative error of iqEPT result at $P=2000$ for $N=8$ is about $7.5\times 10^{-5}$. For $N=12$, it is about $10^{-3}$. See Table.\ref{table:energy} for details. It is obvious that the relative error in iqEPT scales much more slowly with respect to $N$ than that in DMRG for a spin ladder. 

One last interesting observation is that the ladders with OBC in the rung are more computationally demanding in iqEPT. For instance, to obtain the same relative error $3.0\times 10^{-5}$ for $N=6$, $P=900$ for OBC while $P=560$ for PBC. By contrast, DMRG favors OBC\cite{Schollwoeck2005}. This difference has two-fold meanings. First, the entanglement entropy of a rung in the ladder with OBC (solid circles in Fig.\ref{fig:entanglement_coupled}(c)) is greater than that with PBC (Fig.\ref{fig:entanglement_coupled}(b)) for each $N$ in comparison. It determines that OBC in the rung is necessarily more challenging to simulate. Second, DMRG's difficulty with PBC is caused by the winding MPS form\cite{Eisert2010} (Fig.\ref{fig:TNSandMPS}(b)).

\subsection{\label{subsec:coupled} GS properties of target model}
\begin{figure}
	\begin{center}
		$\begin{array}{c}
		\mbox{(a)}\\
		\includegraphics[width=18.pc]{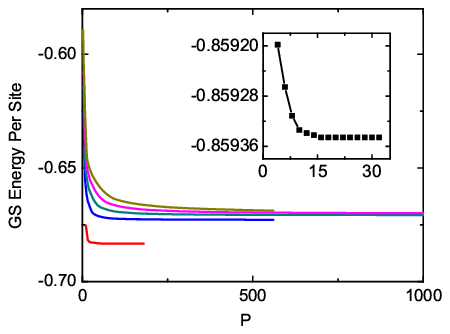}\\
		\mbox{(b)}\\
		\includegraphics[width=18.pc]{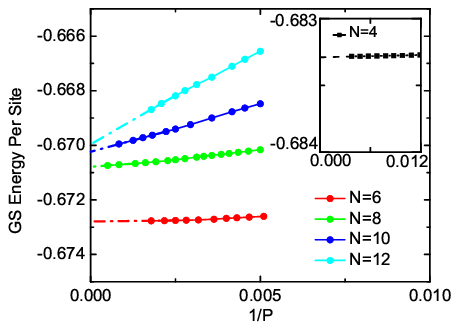}\\   
		\mbox{(c)}\\
		\includegraphics[width=18.pc]{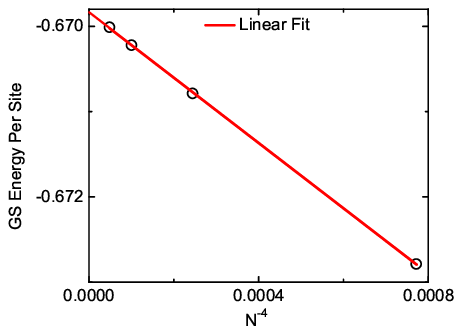}   		
		\end{array}$
		\caption{\label{fig:energy} (a) The convergence of GS energy per site with respect to the MPS rank $P$, for $N=4, 6, 8, 10$ and $12$ from bottom to top. The inset is for $N=2$ in a distinct energy scale. (b) The convergence of energy with respect to $1/P$. The inset is for $N=4$ in a distinct energy scale. They give extrapolations used in (c). (c) The GS energy per site approaches the thermodynamic value as a fourth order function of $\frac{1}{N}\rightarrow 0$, one order faster than the approach from $N$-by-$N$ lattices. Only data for $N=6, 8, 10$ and $12$ are shown due to the very fast decay of $N^{-4}$.}
	\end{center}
\end{figure} 
Our target model is the isotropically coupled ladder of even $N$'s, with PBC imposed in both directions.

One of our main results is the GS energy per site for $N$ up to $14$, shown in Fig.\ref{fig:energy}. In (a), the varying of energy with respect to the MPS rank $P$ is hardly noticeable for a large $P$ value in the energy scale of $0.1$ shown for $N$ up to $12$. Recall that Fig.\ref{fig:decoupled} shows a power relationship between the relative error and $P$ and shows negative linear coefficients $b_1$ defined in equation \eqref{eq:fitting1}. Therefore, the straight lines of energy versus $1/P$ should give reliable extrapolations when $1/P \rightarrow 0$, which is a simpler alternative to the extrapolating process in equation \eqref{eq:fitting}. Fig.\ref{fig:energy}(b) shows energy versus $1/P$ in the scale of $0.01$ for $N=6, 8, 10$ and $12$ from bottom to top, while in the scale of $0.001$ for $N=4$ in the inset. Indeed, the straight lines steadily approach the extrapolations. Table \ref{table:energy} lists the largest MPS rank $P^{\prime}$ used for each $N$'s, the simulated GS energy per site $\epsilon_{P^{\prime}}$, the extrapolated energy $\bar{\epsilon}$ and the relative error $\Delta$. Note that, the kept digits $-0.8593457\left(1\right)$ when $P\ge 16$ for $N=2$ is much more than other $N^{\prime}$s. We plot the extrapolated energies for each $N$ versus $N^{-4}$ in Fig.\ref{fig:energy}(c). Only data for $N=6, 8, 10$ and $12$ are shown due to the very fast decay of $N^{-4}$. They quickly approach the thermodynamic limit value of $-0.66984$ with an uncertainty of $9.6\times 10^{-6}$. Our value agrees well with the accepted values such as $-0.6696\pm0.0003$ by series expansions\cite{Singh1989} and $-0.6693(1)$ by the cluster algorithm\cite{Wiese1994}. It can be compared with the DMRG result of $-0.6768$\cite{Ramos2014}. It is worth mentioning that the finite-size effect fades away in our work by one order of $1/N$ faster than when approaching from $N$-by-$N$ lattice. The energy for an infinity-by-N ($N=12$) lattice has a $2.5\times 10^{-4}$ difference relative to the thermodynamic limit value, as close as that for a $22 \times 22$ lattice (interpolation from Fig. 5 of \cite{Manousakis1991}). 
\begin{table}
	\begin{center}
		\begin{tabular}{ccccc}
			\hline
			N   & $P^{\prime}$  & $\epsilon_{P^{\prime}}$ &$\bar{\epsilon}$ &$\Delta$ \\
			\hline
			2  & $32$ & $-0.85935$ & $-0.85935$ &  \\			
			4  & $266$ & $-0.68328$ & $-0.68329$ & $1.5\times 10^{-5}$\\
			6  & $560$ & $-0.67277$ & $-0.67279$ & $3.0\times 10^{-5}$\\
			8  & $2000$& $-0.67074$ & $-0.67078$ & $6.0\times 10^{-5}$\\
			10 & $1400$& $-0.66996$ & $-0.67017$ & $3.1\times 10^{-4}$\\
			12 & $560$ & $-0.66871$ & $-0.67001$ & $1.9\times 10^{-3}$\\
			14 & $350$ & $-0.66636$ & $-0.66993^*$ & $5.3\times 10^{-3}$\\			
			\hline    
		\end{tabular}
		\caption{\label{table:energy}The simulated GS energy per site $\epsilon_{P^{\prime}}$ at the largest MPS rank tried $P^{\prime}$, the extrapolated energy $\bar{\epsilon}$ and the relative error $\Delta$ for various $N$'s of an infinity-by-$N$ lattice with periodic BC imposed in both directions.$^*$ the extrapolation for $N=14$ is replaced by the interpolation in Fig.\ref{fig:energy}(c).}
	\end{center}
\end{table}
\begin{figure}
	\begin{center}
		$\begin{array}{c}
		\mbox{(a)}	\\
		\includegraphics[width=18.5pc]{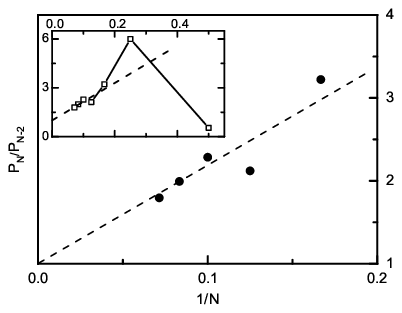}\\
		\mbox{(b)}\\
		\includegraphics[width=18.5pc]{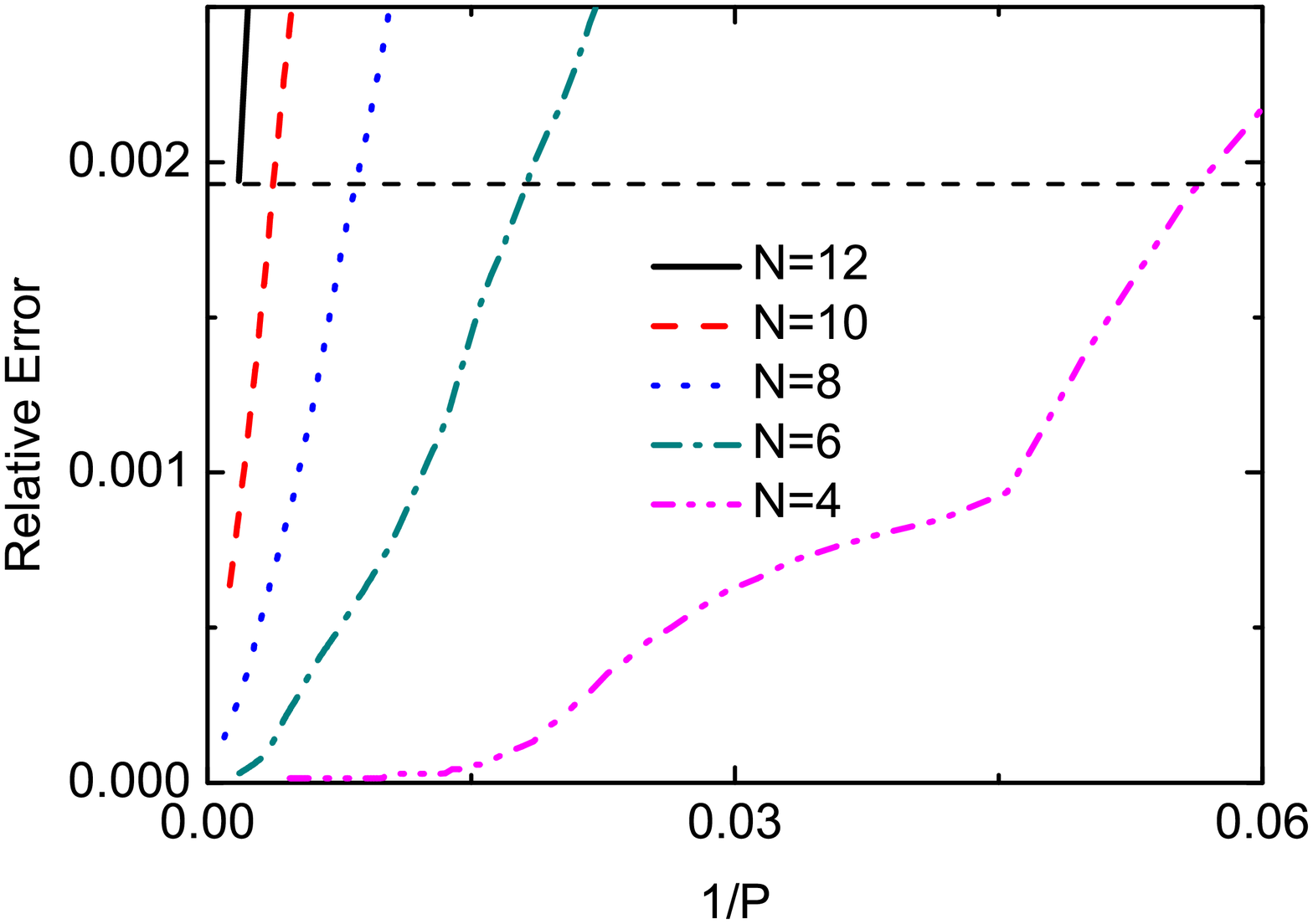}\\
		\end{array}$
		\caption{\label{fig:entanglement} (a) Ratio of the MPS rank $P_N$ (for $N$) to $P_{N-2}$ (for $N-2$), with respect to $\frac{1}{N}$. At both $P_N$ and $P_{N-2}$ the same accuracy is obtained. The tendency of $\frac{P_N}{P_{N-2}}\rightarrow 1$ for $\frac{1}{N}\rightarrow 0$ implies the saturating MPS rank with increasing $N$. The inset shows a larger scale starting from $\frac{P_2}{P_1}$. (b) The horizontal dashed line intercepts the curves of 'relative error versus $1/P$', giving $P_N$'s used in (a), given certain relative error, say $1.9\times 10^{-3}$. The comparison between $N=12$ and $14$ is made for the relative error of $5.3 \times 10^{-3}$, where $P=195$ for $N=12$ and $P=350$ for $N=14$, respectively.}
	\end{center}
\end{figure} 
\begin{figure}
	\begin{center}
		$\begin{array}{c}
		\mbox{(a)}\\
		\includegraphics[width=18.5pc]{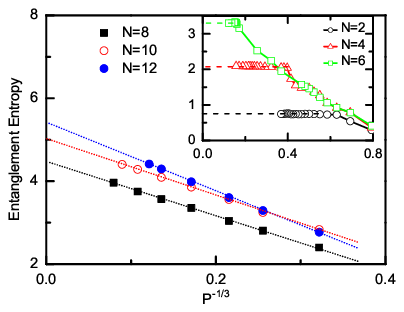}\\
		\mbox{(b)}\\
		\includegraphics[width=18.5pc]{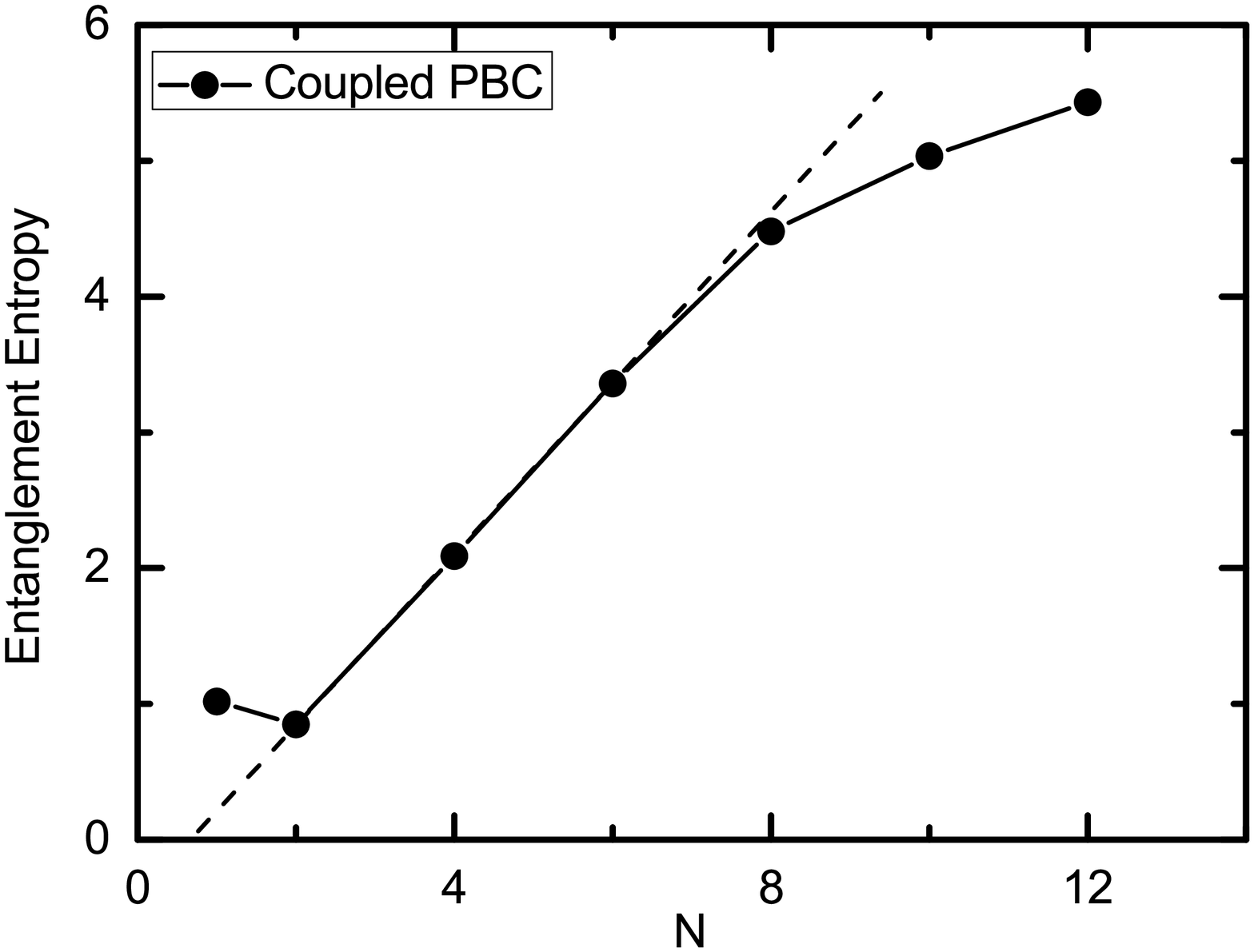}\\		
		\end{array}$
		\caption{\label{fig:entanglement_coupled} Entanglement entropy of an effective site for coupled ladders with PBC in the rung. (a) Entanglement entropy versus $P^{-1/3}$. The steady tendency to $P^{-1/3}=0$ gives extrapolation for $N=8, 10$ and $12$ shown as rectangles, circles and triangles, respectively. Those of $N=2, 4$ and $6$ from bottom to top in the inset exhibit sudden convergence. (b) Entanglement entropy versus $N$. The dashed guiding line shows that the curve are three-segmented. The second segment is linear, covering $N=2,4$ and $6$. The third segment starts to bend when $N\ge 8$ and will saturate when $N\rightarrow\infty$.}
	\end{center}
\end{figure} 

Meanwhile, the most intriguing information from the energy observation is the plot in Fig.\ref{fig:entanglement}(a). It shows the ratio of the MPS rank for a given accuracy, say, $1.9\times 10^{-3}$ for $N$ to that for $N-2$, with respect to $1/N$. Fig.\ref{fig:entanglement}(b) explains how to get each $P_N$. In (a), the dashed guiding line shows the tendency that $P_N/P_{N-2} = 1$ when $1/N\rightarrow 0$. It implies that the increase of entanglement in a MPS wave function built on an effective 1D lattice, whose site is converted from the $N$ sites in the rung of an infinity-by-$N$ lattice, will slow down with $N$ and possibly will be saturated for larger $N$. 

The mechanism of this saturation of $P$ with $N$ is accounted for by the saturating entanglement entropy of an effective site, shown in Fig.\ref{fig:entanglement_coupled}(b). It is now clear that, treating an infinity-by-$N$ lattice as if in 1D does bypass the area law of entanglement entropy for the strongly correlated 2D quantum system only if larger $N$ can be reached. Fig.\ref{fig:entanglement_coupled}(a) shows that the computed entanglement entropy has a linear dependence on $P^{-1/3}$. For the ladders of $N=2, 4$ and $6$, it does suddenly converge\cite{Eisert2010} when $P$ reaches a threshold, forming plateaus shown in the inset. Recall that the definitely gapless decoupled ladder shows continuous convergence of entanglement entropy with respect to the MPS rank in Fig.\ref{fig:entanglement_decoupled}(a). Now that the ladders of $N=8,10$, and $12$ show no plateau either, it is necessary to check whether the sudden convergence of a gapped ladder is not reached yet or the ladder is gapless. These two possibilities shall be explored with more physical quantities. At the moment, however, an immediate assertion can be made that, starting from $N=8$ the lattice is out of the applicable regime of NLSM's prediction that the ladder has a gap which exponentially decays with increasing width. Otherwise, the entanglement entropy shall be linear with all $N$'s.    

We now show that the ladder is still gapped for $N=8$ and that it is ordered hence gapless for $N\ge 10$. We study $C_r$ versus $r$, where $C_r$ is spin-spin correlation at separation $r$ in LD.  Hereafter, we discuss the absolute value of the correlations, despite that they have alternating signs due to the antiferromagnetism. 
\begin{figure}
	\begin{center}
		$\begin{array}{cc}
		\mbox{(a)}&\\
		&\includegraphics[width=18pc]{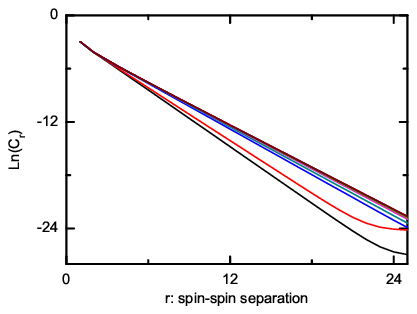}\\
		\mbox{(b)}&\\
		&\includegraphics[width=18pc]{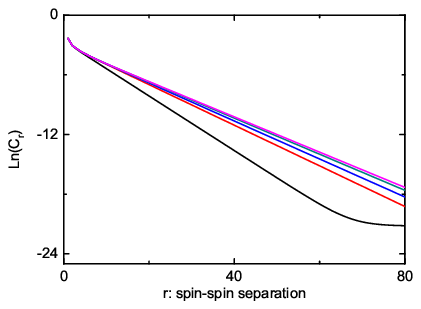}\\
		\mbox{(c)}&\\
		&\includegraphics[width=18pc]{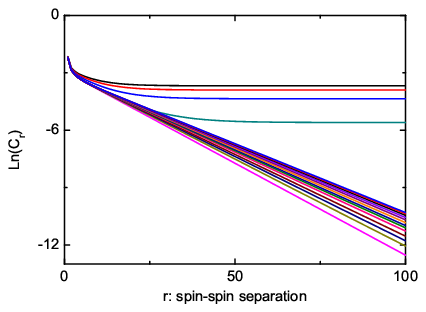}
		\end{array}$
		\caption{\label{fig:correlation} $LnC_r$ versus $r$. The linear tilted lines in logarithmic scale suggest the exponential decay of correlations with respect to separations. They approach the fixed one from bottom when $P=4$ to top when $P=28$ (every augment of $4$ for $P$) for $N=2$ in (a); from bottom when $P=40$ to top when $P=200$ (every $40$) for $N=4$ in (b). However, the beginning lines are flat at top in (c) for $N=6$, starting from top when $P=190$ to the last flat one in the middle zone when $P=250$ (every $20$). It suddenly jumps down to the tilted line at the bottom when $P=270$ and approaches the fixed tilted line when $P=560$ (every $20$).}  
	\end{center}
\end{figure}  
\begin{figure}
	\begin{center}
		$\begin{array}{cc}			
	\mbox{(a)}&\\
		& \includegraphics[width=18pc]{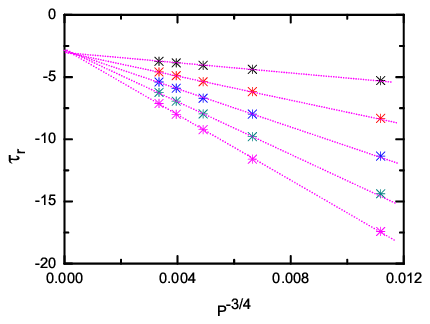}\\
	\mbox{(b)}&\\
		& \includegraphics[width=18pc]{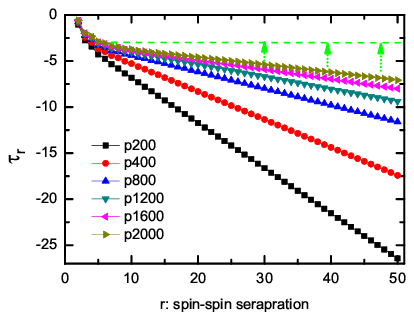}\\
	\mbox{(c)}&\\
		& \includegraphics[width=18pc]{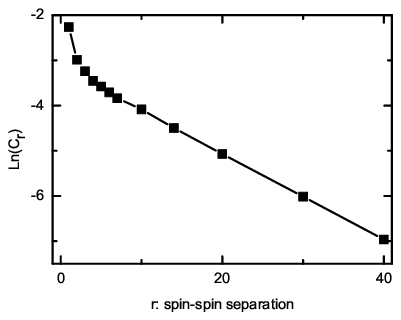}\\		
		\end{array}$
		\caption{\label{fig:N8C} $C_r$ and $\tau_r\equiv Ln\left(LnC_r-LnC_{r+1}\right)$ for $N=8$. (a) $\tau_r$ versus $P^{-3/4}$ at separations $r=10, 20, 30, 40$ and $50$ from top to bottom. (b) $\tau_r$ versus $r$ at various $P$'s. They asymptotically approach the dashed curve obtained by the extrapolation in (a). Trace along the asymptotic curve in (c), starting from $LnC_1=Ln\frac{\bar{\epsilon}_0}{6}$, yields the asymptotic curve of $LnC_r$ versus $r$.}
	\end{center}
\end{figure} 
\begin{figure}
	\begin{center}
		$\begin{array}{cc}
		\mbox{(a)}&\\
		& \includegraphics[width=18pc]{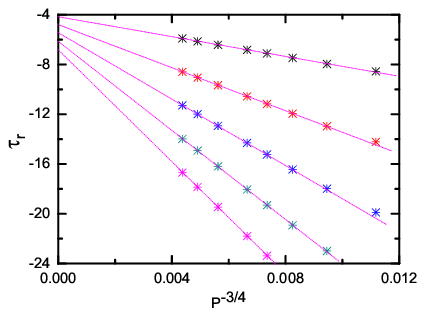}\\
		\mbox{(b)}&\\	
		& \includegraphics[width=18pc]{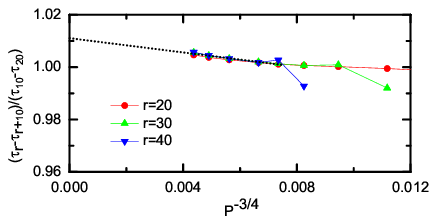}\\
		\mbox{(c)}&\\	
		& \includegraphics[width=18pc]{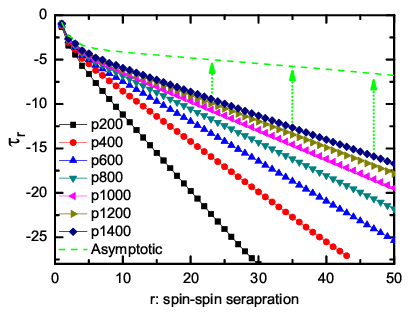}\\
		\mbox{(d)}&\\	
		& \includegraphics[width=18pc]{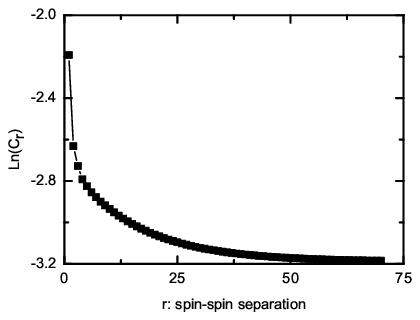}			
		\end{array}$
		\caption{\label{fig:N10C} $C_r$ and $\tau_r\equiv Ln\left(LnC_r-LnC_{r+1}\right)$ for $N=10$. (a) $\tau_r$ versus $P^{-3/4}$ at $r=10, 20, 30, 40$ and $50$ from top to bottom. (b) $\left(\tau_r-\tau_{r+10}\right)/\left(\tau_{10}-\tau_{20}\right)$ versus $P^{-3/4}$ at $r=20, 30$ and $40$. (c) $\tau_r$ versus $r$ at various $P$'s. They asymptotically approach the dashed curve obtained by the extrapolation in (a). (d) Trace along the asymptotic curve in (c) yields the asymptotic curve of $LnC_r$ versus $r$.}
	\end{center}
\end{figure}

Fig.\ref{fig:correlation} is shown in the semi-logarithmic scale for $N=2, 4$ and $6$. The straight tilted lines indicate exponential decays with respect to the spin-spin separation. Comparison of the spin-spin separation needed for the same value of spin-spin correlation gives the ratio of correlation lengths for these three lattices. It is $1:4:9$. It is worth noting the behavior of $N=6$. It looks straight when $P\le 250$ but then jumps down to the bottom when $P=270$, and finally converges to the fixed line. It is a clear indication of the competition between order and disorder. 

For $N\ge 8$, we didn't obtain the converged plot of $C_r$ versus $r$ due to the larger entanglement entropy. For them, we study a new quantity $\tau_r\equiv Ln\left(LnC_r-LnC_{r+1}\right)$. It is the varying rate of $LnC_r$. If this rate is a negative constant, the correlation decays exponentially with $r$. If the rate somehow decays with $r$, the correlation is a constant at infinite separation. Hence, the lattice is ordered. Fig.\ref{fig:N8C}(b) and Fig.\ref{fig:N10C}(b) both show that this quantity is linear with $r$ at various MPS rank $P$'s, as explained in Sec.\ref{sec:correlation}. But their asymptotic ($P\rightarrow \infty$) behaviors are different. For $N=8$, $\tau_r$ asymptotically becomes a negative constant for large $r$'s shown as the dashed curve in Fig.\ref{fig:N8C}(b). This negative constant is obtained when the lines of $\tau_r$ versus $P^{-3/4}$ for various large $r$'s converge to the same value when $p\rightarrow \infty$ in (a). Starting from $C_1=\bar{\epsilon}_0/6$ and then tracing along the asymptotic curve in (b), we obtain the dependence of $LnC_r$ on $r$ in (c). It is seen that, $C_r$ for $N=8$ decays exponentially with $r$. Nevertheless, For $N=10$, Fig.\ref{fig:N10C}(a) shows that the lines of $\tau_r$ versus $P^{-3/4}$ don't converge to the same value when $p\rightarrow \infty$. (b) further shows that $\tau_{r}-\tau_{r+10}$ are equal for $r=10, 20, 30$ and $40$, implying that the lines in (a) are equally spaced. Thus, the dashed asymptotic curve in (c) has a constant negative slope for large $r$'s. Tracing along the asymptotic curve, we obtain the dependence of $LnC_r$ on $r$ in (d). $LnC_r$ hence $C_r$ becomes a nonzero constant at infinite spin-spin separation. The ladder of $N\ge 10$ is ordered. Since no external pinning magnetic field\cite{White2007} is applied, it implies that the spin rotational symmetry is spontaneously broken. Our finding, that lattice of $N\le 6$ is not ordered, is fully consistent with the previous report\cite{Greven1996} that a gap exist for lattices of $300 \times N$, $N\le6$. The gap leads to the fast exponential-like decay for spin-spin correlations reported for those lattices. Nevertheless, for the first time we show with strong numerical evidence that the spin rotational symmetry spontaneously breaks for a spin-$\frac{1}{2}$ lattice of $N\ge 10$. Since a spontaneously ordered GS is regarded as a 2D characteristic by the existing theories, such as SWT, NLSM and Mermin-Wager theory\cite{Mermin1966,Coleman1973}, the spontaneous symmetry breaking defines a quantum dimensional transition from 1D including quasi-1D to 2D at a finite ladder width $N$.

\subsection{\label{subsec:space-reduction} effects of space reduction in matrix product state}
\begin{figure}
	\begin{center}
		$\begin{array}{cc}
		\mbox{(a)}&\\
		& \includegraphics[width=18pc]{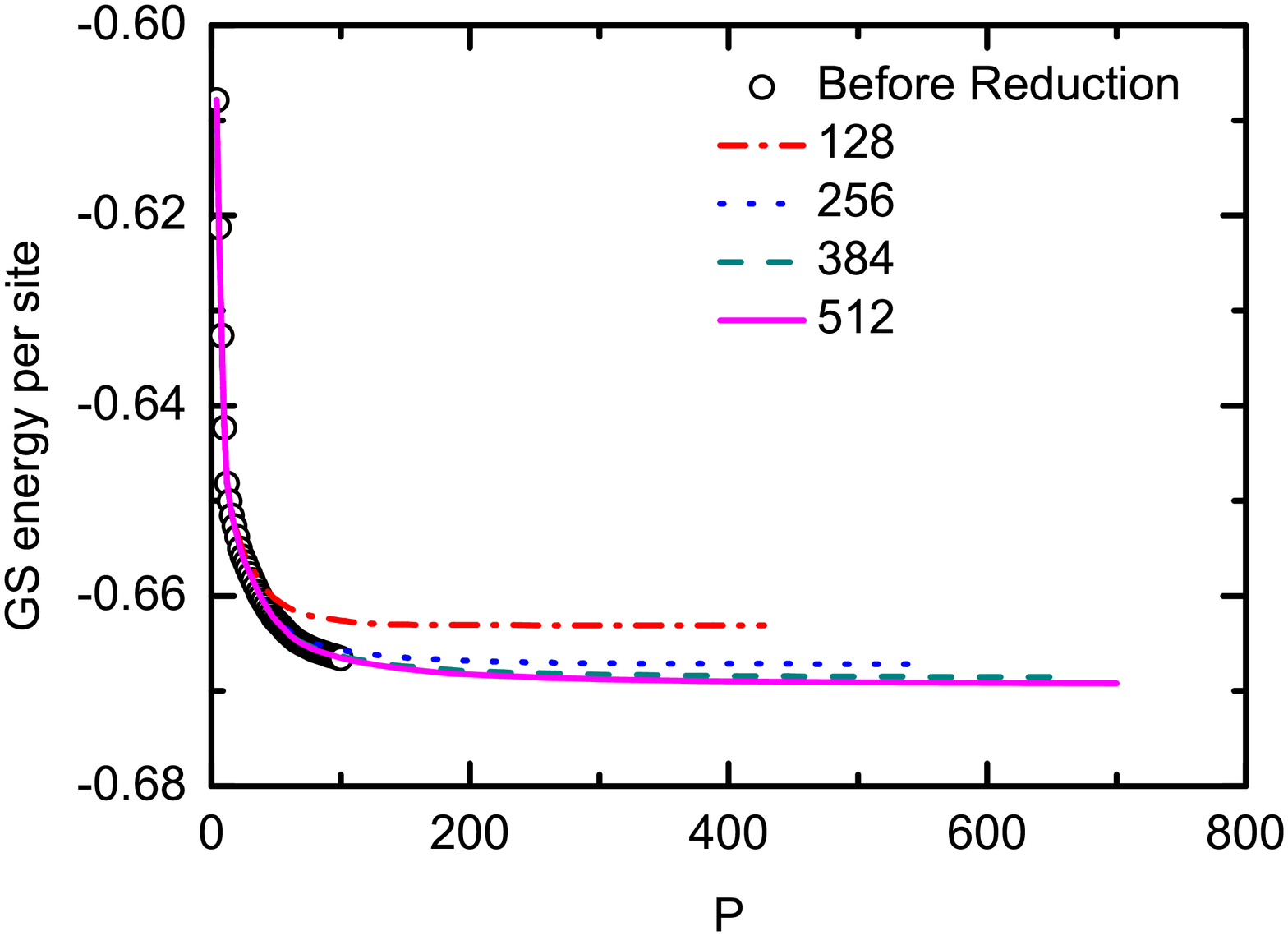}\\
		\mbox{(b)}&\\	
		& \includegraphics[width=18pc]{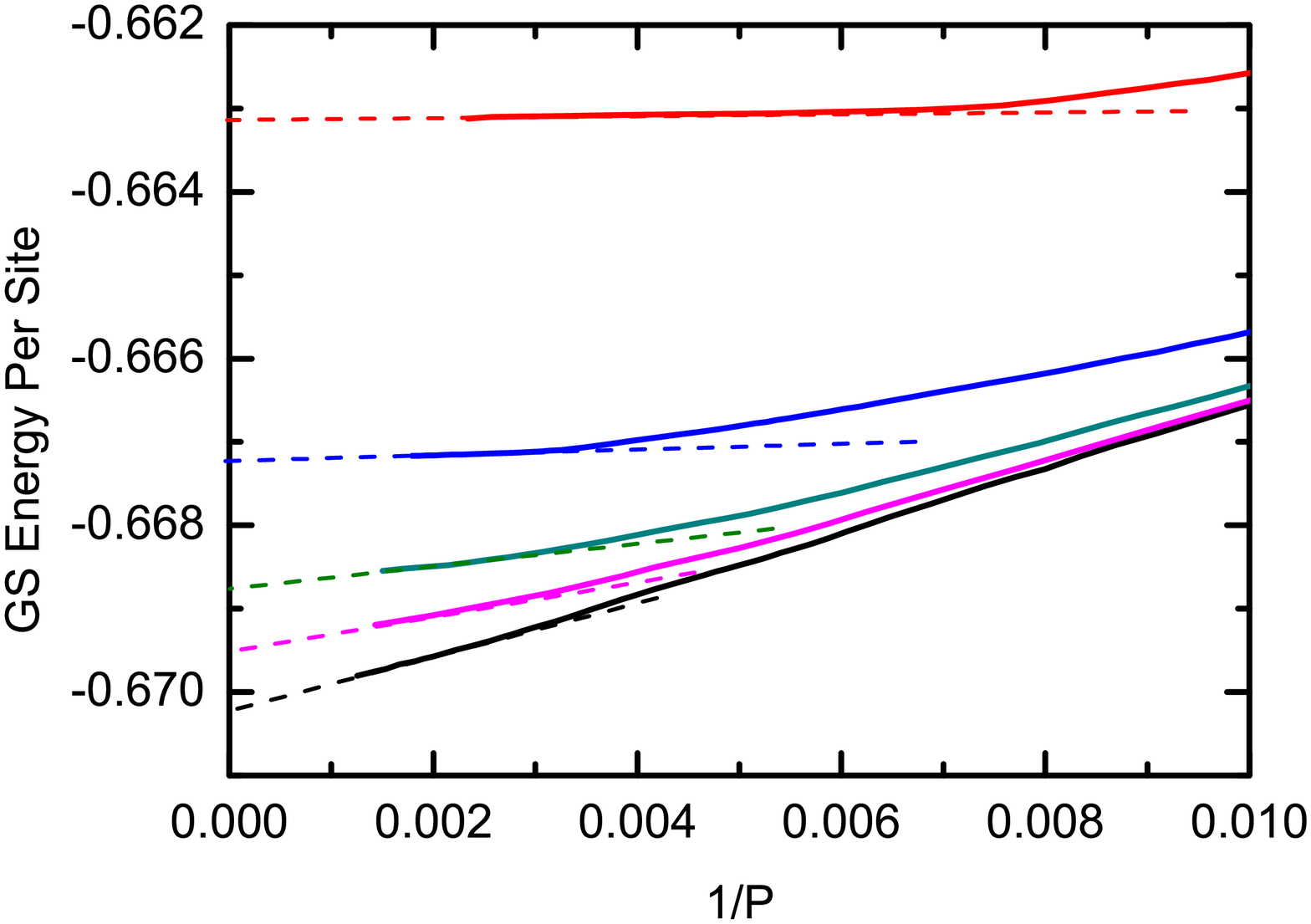}\\
		\mbox{(c)}&\\	
		& \includegraphics[width=18pc]{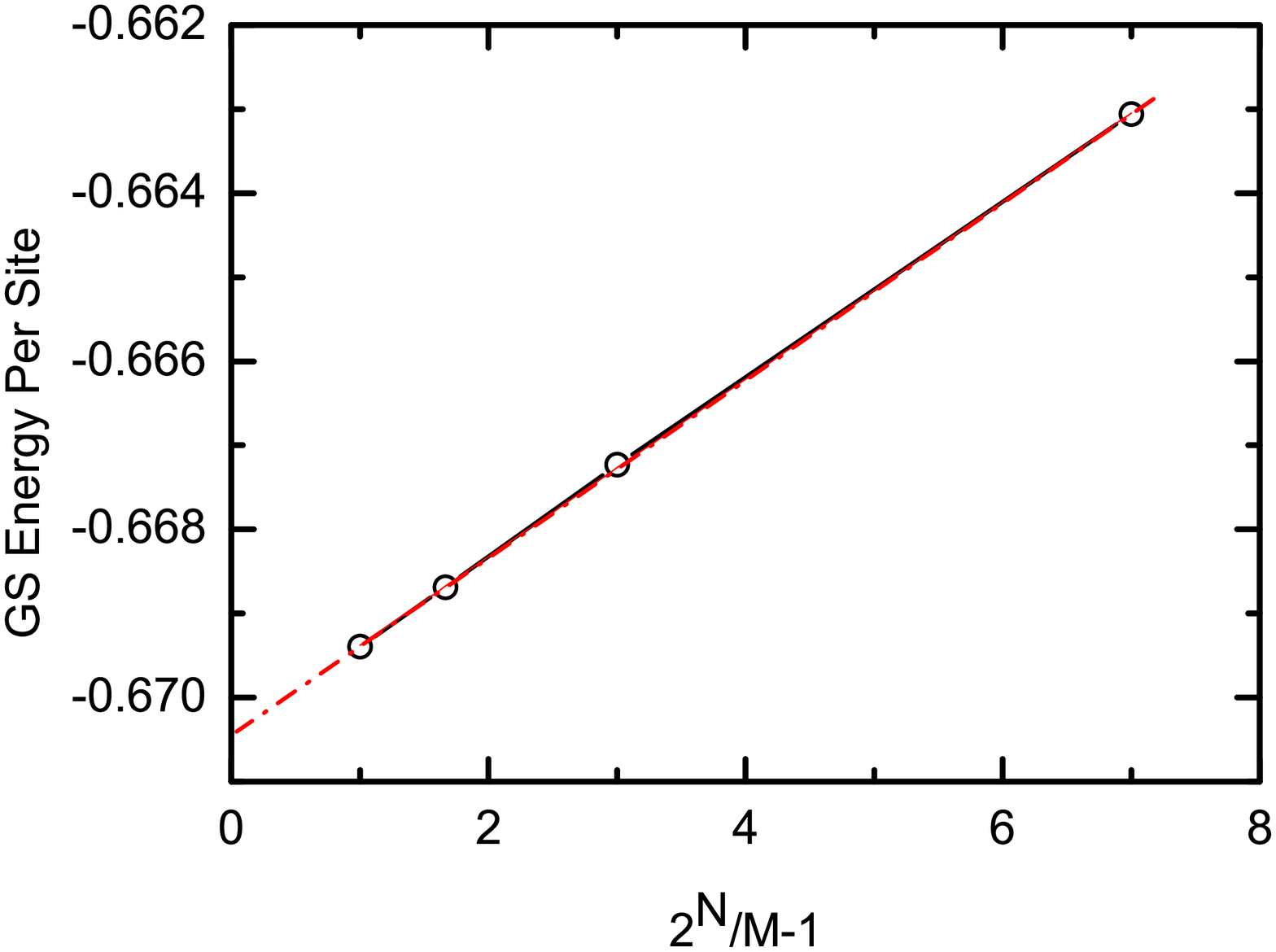}		
		\end{array}$
		\caption{\label{fig:energy1}Effect of space reduction in MPS for $N=10$. (a) GS energy per site versus MPS rank $P$. Open circles denote the solution before reduction. Result at $P_1=100$ is used to reduce the space rank to $128$, $256$, $384$ and $512$, yielding new solutions shown as dot-dashed, dotted, dashed and solid curves. (b) GS energy versus $1/P$. Tangents of convergence yield the extrapolated energies for various space ranks, $128$, $256$, $384$, $512$ and $1024$ (unreduced) from top to bottom. (c) Extrapolation of the energy in unreduced spaces using those obtained in reduced spaces.}
	\end{center}
\end{figure} 
\begin{figure}
	\begin{center}
		$\begin{array}{c}
		\includegraphics[width=18.pc]{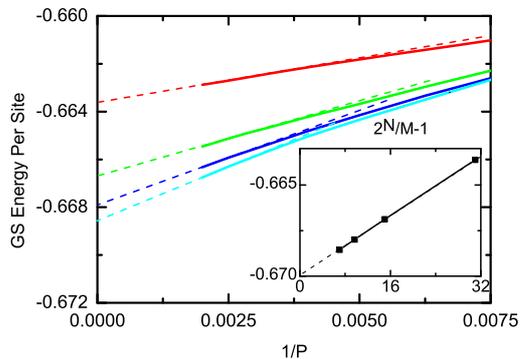}\\
		\end{array}$
		\caption{\label{fig:energy2}Use of space reduction in MPS for $N=14$. Tangents of the convergence of GS energy per site versus $1/P$ yield extrapolated energies for various space ranks, $512$, $1024$, $1536$ and $2048$ from top to bottom. The inset extrapolates GS energy per site in the unreduced space using those extrapolations obtained in reduced spaces.}
	\end{center}
\end{figure} 
\begin{figure}
	\begin{center}
		$\begin{array}{cc}			
		\mbox{(a)}&\\
		& \includegraphics[width=18.pc]{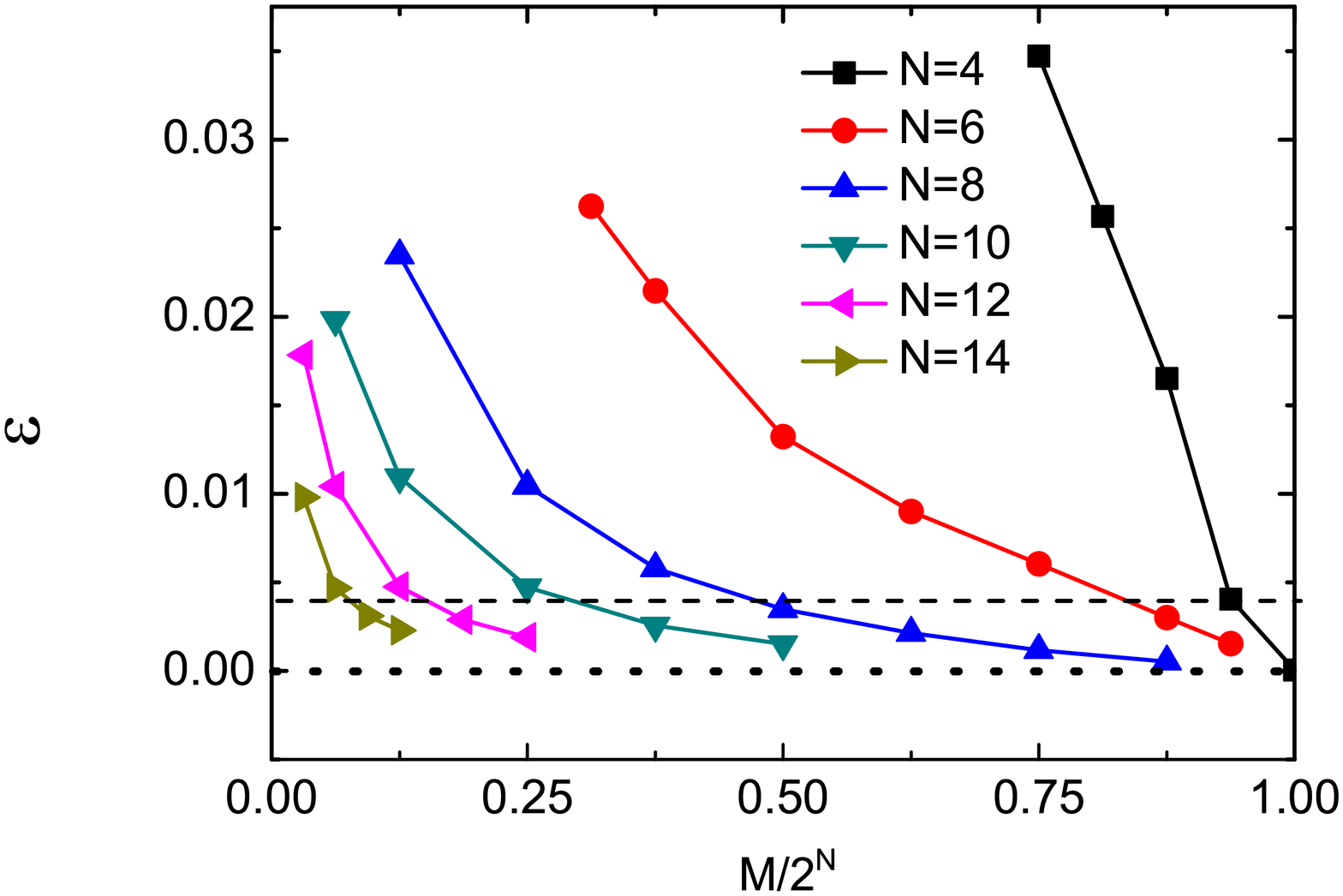}\\
		\mbox{(b)}&\\	
		& \includegraphics[width=18.pc]{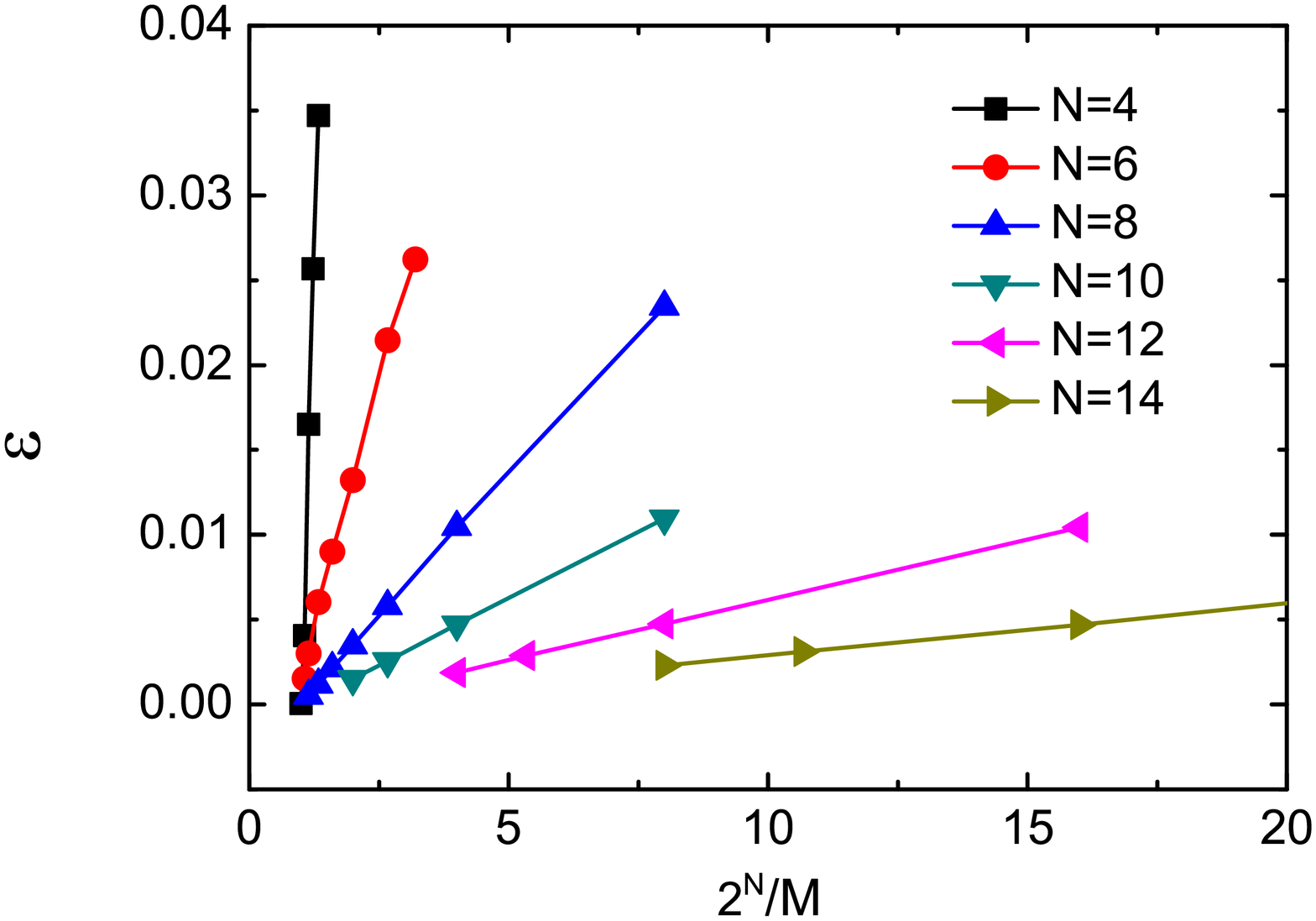}\\
		\end{array}$
		\caption{\label{fig:spaceerror} Relative error versus (a) $M/2^N$ (reduction ratio) and (b) $2^N/M$ (inverse ratio). In (a), The dotted line gives a reference of zero error, while the dashed line intercepts each curve to give the reduction ratio at a certain accuracy. (b) Energies obtained in reduced spaces for lattices of larger $N$ approach more linearly to those in unreduced space. Fig.\ref{fig:energy1}(c) and Fig.\ref{fig:energy2} show such examples for $N=10$ and $14$, respectively.}
	\end{center}
\end{figure} 
\begin{figure}
	\begin{center}
		$\begin{array}{ccc}			
		&\includegraphics[width=18.pc]{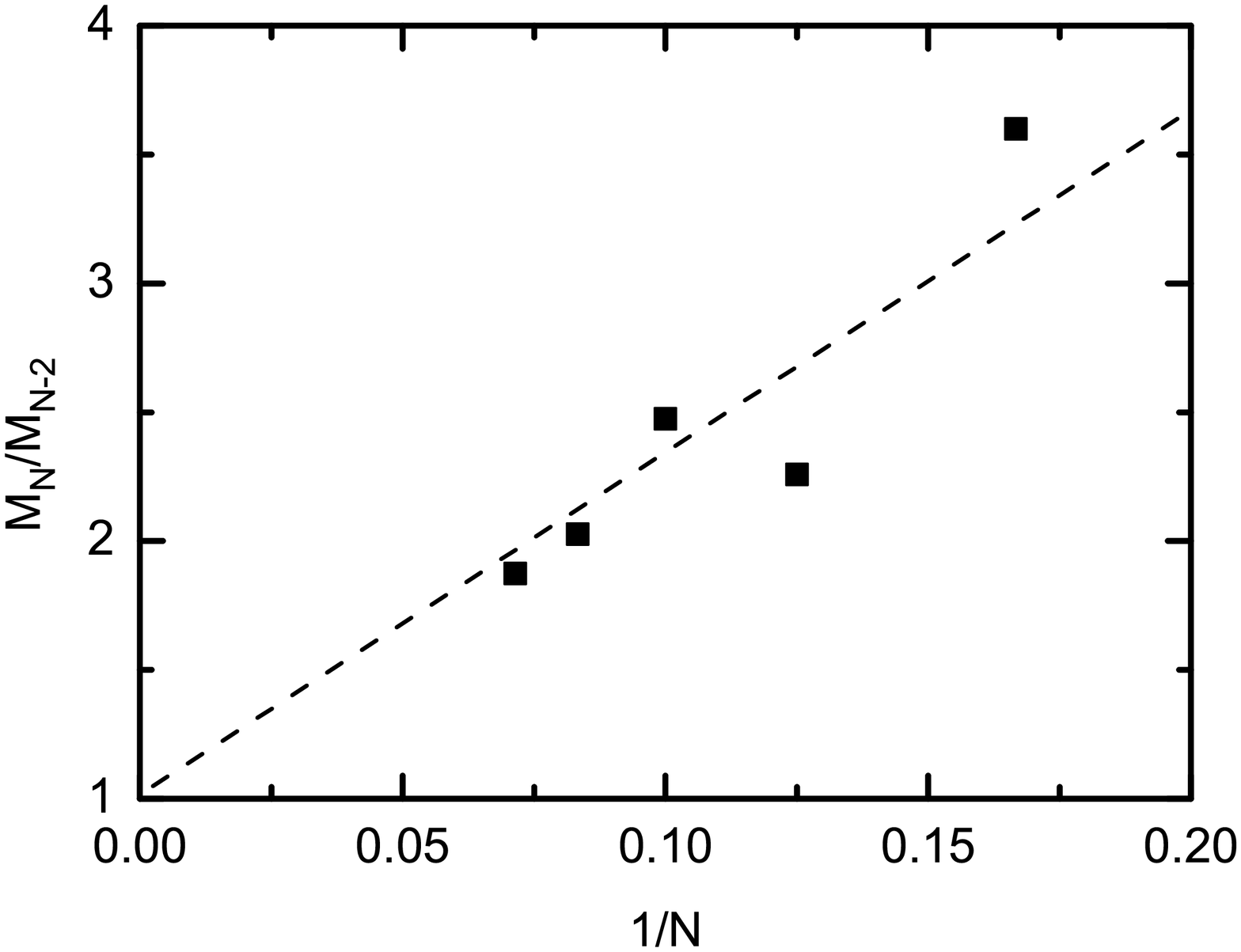}& \\
		\end{array}$
		\caption{\label{fig:fixedpoint} $M_N/M_{N-2}$ (ratio of numbers of basis vectors kept for $N$ and $N-2$, respectively) versus $1/N$. The linear fit overlaps with the guiding dashed line to $1$ when $N\rightarrow \infty$.}
	\end{center}
\end{figure} 
The effect of space reduction in MPS is shown with the example of $N=10$ in Fig.\ref{fig:energy1}. In (a), simulated data in the original space of rank of $2^{10}$ are shown as open circles. At $P_1=100$, the solution is used to reduce the space rank to $128$, $256$, $384$ and $512$ to yield solutions in dot-dashed, dotted, dashed, and solid curves, respectively. Except the reduced rank $128$, simulations for other reductions reproduce the solution before reduction when $P\le P_1$. The closing gaps between flattening curves are confirmed in (b), where the energies versus $1/P$ are plotted for space ranks $128$, $256$, $384$ and $512$ from top to bottom. The simulation in original space is also carried on after $P_1$, shown as the bottom curve in the same plot. All curves show convergency. The extrapolation by tangents of those converging curves yields energies in spaces of both various reduced sizes and the original size. Those in the reduced spaces are used to extrapolate the energy in the unreduced space, as shown in (c). There, the linear fit yields $-0.6704$, agreeing well with $-0.67022$ by extrapolation using the data obtained before space reduction in (b). Note that this scheme which extrapolates the result in the original space with the data obtained in reduced spaces, is much more computationally efficient so as to allow simulation at larger $P$ values. 

We run simulations for $N=14$ in various reduced spaces of ranks $512$, $1024$, $1536$ and $2048$ up to $P=500$, shown from top to bottom in Fig.\ref{fig:energy2}. The lowest energy without extrapolation is $-0.66676$ at $P=500$ in the reduced space of size $2048$, lower than the previously reported value of $-0.66636$ at $P=350$ that is the largest $P$ value handleable in the unreduced spaces. Meanwhile, the inset extrapolates to $-0.66998$ with the difference of $5 \times 10^{-5}$ from $-0.66993$ which was obtained by the interpolation in Fig.\ref{fig:energy}(c).

Fig.\ref{fig:spaceerror}(a) shows the result of $\epsilon$ versus $M/2^N$ for various $M^{\prime}s$ and $N^{\prime}s$, where $2^N$ is the original space rank; $\epsilon$ is the relative error between the energies obtained before and after reduction. It is seen that only $1/8$ of the original space size $2^{12}$ is needed for $N=12$, to achieve a relative error of $4.1\times 10^{-3}$. In comparison, the same accuracy for $N=4$ is obtained with $15$ out of $2^4$ basis vectors. For $N=2$, no reduction will achieve good accuracy. Fig.\ref{fig:spaceerror}(b) shows the dependence of relative error on $2^N/M$. Larger lattices ($N\ge 8$) show a linear dependence, which is a reconfirmation for the reliability of extrapolating results using simulation in reduced spaces. Fig.\ref{fig:energy}(b) and (c) illustrate such an example for a lattice of $N=10$. Fig.\ref{fig:energy2} shows another example for $N=14$.

We plot in Fig.\ref{fig:fixedpoint} $M_N/M_{N-2}$ (ratio of numbers of basis vectors kept to achieve the same accuracy for $N$ and $N-2$, respectively) versus $1/N$. It shows that this ratio tends to approach $1$ when $N\rightarrow \infty$. As discussed in Sec.\ref{subsec:mpsept}, a saturating number of significant diagonal density matrix element of an effective is responsible for the saturating entanglement entropy hence for the saturating MPS rank $P$, when $N$ increases. Fig.\ref{fig:entanglement}(b) and Fig.\ref{fig:fixedpoint} are indeed consistent.

\section{\label{sec:Conclusion}Conclusion}
In conclusion, the way we treated the infinity-by-$N$ quantum lattice as 1D effective lattice, converting $N$ lattices in the rung into an effective site, enables us to handle the unprecedented lattice size with $N$ up to $14$. We show that both the number of significant diagonal density matrix elements and the entanglement entropy of an effective site saturate with increasing $N$. The former is responsible for the latter. It bypasses the area law of entanglement entropy for the 2D quantum lattice. Our results for such a lattice with OBC in the rung are progressively more accurate for larger $N$'s than DMRG.

For the target model with PBC in both rung and LD, NLSM's prediction, that the lattice will have a gap which exponentially decays with $N$ till $N\rightarrow\infty$, is shown to only fully apply to $N \le 6$ and partially apply to $N=8$ whose gap does not exponentially decay. By contrast, our data revealed the quantum dimensional transition from 1D (including quasi-1D) to 2D that takes place at a critical width $N=10$, with emerging\cite{Anderson1972} order parameters. 

At last, it is worth comparing our observation of quantum dimensional transition with the assertion of Mermin-Wagner theory\cite{Mermin1966,Coleman1973}. It states that the Heisenberg model cannot have spontaneous ordering (spin rotational symmetry breaking) at any finite temperature in both 1D and 2D. For such a model, spontaneous symmetry breaking of GS is different. But, despite the possible failure\cite{Vojta2005} of quantum-classical mapping, it is used to show within the framework of Mermin-Wagner theory that Heisenberg model supports spontaneous ordering in 2D but excludes magnetic order in a pure 1D. When an infinity-by-$N$ square lattice is converted into an effective 1D chain, we show that this effective chain may or may not support spontaneous ordering depending on $N$. It is consistent with the previous findings of spontaneous ordering for a 1D chain which is not so pure as to include unequal spins\cite{Tian1997,Kolezhuk1997}.

\section{\label{sec:outlook}Outlook}
The saturating entanglement grantees that the MPS rank, which otherwise exponentially increases with $N$, saturates as well, relieving the major computational burden related to the MPS size. It is instructive to exhaust other factors which will cause an exponential growth of computational burden with respect to $N$ in this method. The first such a factor is the linking complexity in MPO that, however, in this work is reduced to a linear relationship with $N$ by the entanglement perturbation of MPO. The second and also the last such a factor is the exponentially increasing number of local quantum states on the effective site. It is $2^N$ for spin-$\frac{1}{2}$. The limited number of significant diagonal density matrix elements of an effective site enables an efficient reduction of the space in MPS hence eliminates the last exponential factor in this method. It is possible that a 2D infinity-by-infinity quantum lattice physically behaves like a 1D lattice which has limited significant local states on a slice and is linked with limited entanglement between neighboring slices, when looked from any direction of its two dimensions. The method used in this work is a promising numerical tool when studying the 2D strong correlation in this way. 

Meanwhile, the emerging local magnetization in those infinity-by-$N$ lattices with $N\ge 10$ shows different finite-size effect from that of an $N$-by-$N$ or $\alpha N$-by-$N$\cite{White2007} ($\alpha$ is a small integer) lattice. Since no pinning magnetic field $B$ is needed, extrapolating the thermodynamic limit value will be simpler. Staggered magnetization, one of the most fundamental physical quantities for quantum spins, is worthy more investigation along this line.

Note that the space reduction in MPS shown in this study can be readily extended to any form of MPS or TNS based methods such as PEPS, whenever they are built on a blocked quantum system.

\hspace{11cm}

\begin{acknowledgments}
This work was supported by NRF (National Honor Scientist Program 2010-0020414) and KISTI (KSC-2017-C3-0081). We thank D. C. Yang for proofreading and discussion.
\end{acknowledgments}



\appendix 
\section{\label{sec:appendix} Correlations with MPS Quantities}
\begin{figure}
	\begin{center}
		$\begin{array}{cc}			
		\mbox(a)&\\
			&\includegraphics[width=18.5pc]{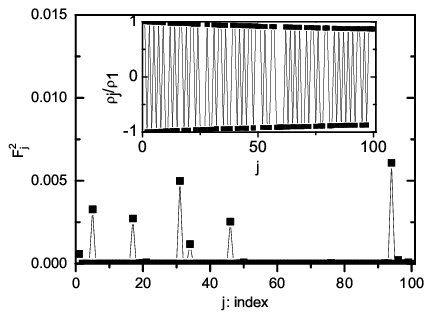} \\
		\mbox(b)&\\
			&\includegraphics[width=18.5pc]{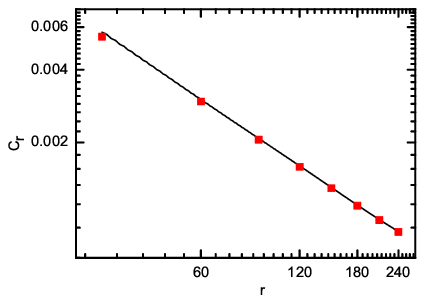} \\	
		\mbox(c)&\\
			&\includegraphics[width=18.5pc]{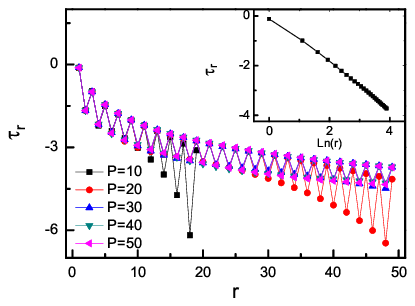} \\		
		\end{array}$
		\caption{\label{fig:spin-chain} QLRO correlations of a spin chain. (a) The largest few eigenvalues, which are shown in inset to be nearly degenerate with $\varrho_1$, have less significant $F_j$'s. They make small contributions to $C_r$ that slowly decay with $r$. The eigenvalues which have significant $F_j$'s are definitely smaller than $\varrho_1$. They make contributions that are large at smaller $r$'s but decay rapidly with $r$. (b) The solid line in the log-log view of $C_r$ versus $r$ collects contributions to $C_r$ from all eigenvalues. Rectangles collect contributions only from those mentioned in (a). They reproduced the power-law decay of QLRO correlations. (c) $\tau_r\equiv Ln\left( LnC_r-LnC_{r+1}\right)$ versus $r$. The even-odd branched curve converges with the MPS rank $P$. The inset takes the odd branch as an example to show $\tau_r$ is linear with $Ln\left(r\right)$, in sharp contrast to the linear dependence of $\tau_r$ on $r$ for both disorder and order.}
	\end{center}
\end{figure} 
\begin{figure}
	\begin{center}
		$\begin{array}{cc}			
		\mbox(a)&\\
			&\includegraphics[width=18.5pc]{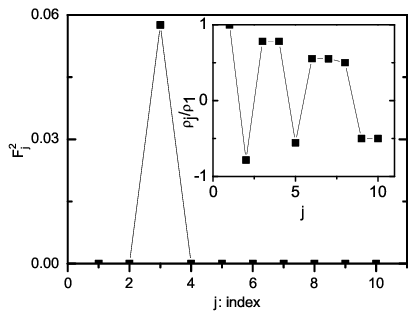} \\
		\mbox(b)&\\
			&\includegraphics[width=18.5pc]{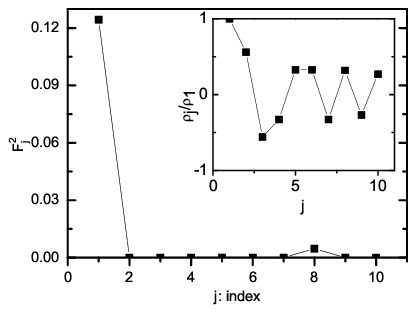} \\	
		\end{array}$
		\caption{\label{fig:disorder-order} Eigenvalue structure of $B$ for (a) the disordered spin ladder of $N=4$ and (b) the ordered ladder of $N=10$.}
	\end{center}
\end{figure} 
Bosonization\cite{Luther1974,Luther1975} predicts a power-law decay of spin-spin correlations $C\left(r\right)=r^{-1}$ for an antiferromagnetic spin-$\frac{1}{2}$ chain, r being the spin-spin separation. So, the spin chain is QLRO. In Sec.\ref{sec:correlation} we discussed in the framework of MPS that the eigenvalue structure of the building unit $B$ (defined in the second equation of \eqref{eq:relabel}) for $\langle g\mid g\rangle$ is responsible for spin-spin correlations, where $\mid g\rangle$ is the GS. 

Fig.\ref{fig:spin-chain}(a) shows $F_j$'s defined in equation \eqref{eq:Fk} for the first $100$ largest eigenvalues of $B$. Only few have significant $F_j$'s. In the inset, the ratio $\varrho_j/\varrho_1$ versus $j$ shows that, a few largest eigenvalues are almost degenerate with $\varrho_1$. We then plot $C_r$ versus $r$ for spin chain in Fig.\ref{fig:spin-chain}(b). The linear solid line in log-log view is obtained according to equation \eqref{eq:correlation2}, showing a power-law decay. This result contains contributions from all eigenvalues of $B$. Nevertheless, the scatters represent data that only collects the contributions from a few largest eigenvalues, that are nearly degenerate with $\varrho_1$ but have less significant $F$'s, and from those that are definitely smaller than $\varrho_1$ but have more significant $F$'s shown as spikes in (a). The slowly varying constant-like correlation-contribution by the former adds up with the fast exponentially decaying correlation-contribution by the latter, reproducing the power-law decay in a large range of $r$. Meanwhile, Fig.\ref{fig:spin-chain}(c) plots $\tau_r$ defined in equation \eqref{eq:tau} versus $r$. The odd and even series both converge with the MPS rank $P$ to nonlinear curves. The inset shows that, for instance, the odd series of $\tau$ versus $Ln\left(r\right)$ is linear. It is in sharp contrast to the linear dependence of $\tau$ on $r$ of either disorder (Fig.\ref{fig:N8C}(b)) or order (Fig.\ref{fig:N10C}(c)). 

Fig.\ref{fig:disorder-order} shows the eigenvalue structure of $B$ for disorder in (a) and order in (b). For the disordered spin ladder of $N=4$ in (a), only the third largest eigenvalue of $B$ has non-vanishing $F_3$ shown as the spike. Inset shows that $\varrho_3/\varrho_1$ is significantly smaller than $1$. It leads to the exponential decay of $C_r$ till zero for large $r$, making $\tau_r$ a constant. If $F_1$ is not converged with the MPS rank $P$ yet and hence does not vanish, it will be a small value compared with $F_3$. $\tau_r$ linearly decreases with $r$. The decreasing rate asymptotically becomes zero when $P\rightarrow \infty$. See Fig.\ref{fig:N8C} for such an example of $N=8$. Meanwhile, for the ordered spin ladder of $N=10$ in Fig.\ref{fig:disorder-order}(b), only the first and eighth largest eigenvalues of $B$ have non-vanishing $F_1$ and $F_8$ shown as the spikes. $\varrho_8/\varrho_1$ is significantly smaller than $1$ in inset. It belongs to Case 1 described in Sec.\ref{sec:correlation}. $LnC_r$ exponentially decays to a nonzero constant when $r$ is large, as Fig.\ref{fig:N10C} shows. Thus, the lattice has a non-vanishing correlation even at infinite separation and hence is ordered.          

\vspace*{10pc}

\bibliography{Heisenberg}

\end{document}